\documentclass[aps,prb,preprint,superscriptaddress, nobibnotes]{revtex4-1}

\usepackage{comment}
\usepackage{graphicx}
\usepackage{amsmath}
\usepackage{amssymb}
\usepackage{bm}
\usepackage{mhchem}
\usepackage{gensymb}

\begin{document}


\title{Band Structure Driven Thermoelectric Response of Topological Semiconductor \ce{ZrTe5}}
\author{Junbo Zhu}
\affiliation{Department of Physics, Massachusetts Institute of Technology, Cambridge, Massachusetts 02139, USA}
\author{Changmin Lee}
\homepage{Present Address: Lawrence Berkeley National Laboratory, Berkeley, California 94720, USA}
\affiliation{Department of Physics, Massachusetts Institute of Technology, Cambridge, Massachusetts 02139, USA}
\author{Fahad Mahmood}
\homepage{Present Address: Department of Physics, University of Illinois at Urbana-Champaign, Urbana, 61801 IL, USA}
\affiliation{Department of Physics, Massachusetts Institute of Technology, Cambridge, Massachusetts 02139, USA}

\author{Takehito Suzuki}
\homepage{Present Address: Department of Science, Toho University, Funabashi City, Chiba 274-8510, Japan}
\affiliation{Department of Physics, Massachusetts Institute of Technology, Cambridge, Massachusetts 02139, USA}
\author{Shiang Fang}
\affiliation{Department of Physics and Astronomy, Center for Materials Theory, Rutgers University, Piscataway, NJ, USA}
\affiliation{Department of Physics, Massachusetts Institute of Technology, Cambridge, Massachusetts 02139, USA}
\author{Nuh Gedik}
\affiliation{Department of Physics, Massachusetts Institute of Technology, Cambridge, Massachusetts 02139, USA}
\author{Joseph G. Checkelsky}
\affiliation{Department of Physics, Massachusetts Institute of Technology, Cambridge, Massachusetts 02139, USA}


\date{\today}

\begin{abstract}
We report a transport, thermodynamic, and spectroscopic study of the recently identified topological semiconductor ZrTe$_5$ with a focus on elucidating the connections between its band structure and unusual thermoelectric properties.  Using time and angle resolved photoemission spectroscopy (tr-ARPES) we observe a small electronic band gap and temperature dependent Fermi level which traverses from a single valence to conduction band with lowering temperature, consistent with previous reports.  This low temperature Fermi surface closely matches that derived from quantum oscillations, suggesting it is reflective of the bulk electronic structure.  The Seebeck and low field Nernst response is characterized by an unusually large and non-monotonic temperature evolution.  We find this can be quantitatively explained using a semiclassical model based on the observed band character and a linear temperature shifting of the Fermi level.  Additionally, we observe a large, non-saturating enhancement of both thermoelectric coefficients in magnetic field.  We show this can be captured by the Zeeman energy associated with a large effective $g$-factor of 25.8 consistent with that derived from Lifshitz-Kosevich analysis of the quantum oscillations.  Together these observations provide a comprehensive picture of ZrTe$_{5}$ as a model high mobility semiconductor and potential platform for significant magnetic field driven thermoelectricity. 


\end{abstract}


\maketitle

\section{Introduction}

The layered material ZrTe$_5$ has seen a renewed interest in recent years following the prediction that it could harbor topologically non-trivial ground states \cite{WengPRX14}.  Originally investigated as a potential charge density wave host \cite{okada1980giant,Okada1982}, the system shows a large, sign-changing Seebeck response long interpreted as evidence for a change of dominant carrier type \cite{Jones1982Sxx}.  The nature of this crossover has become of heightened interest, as different theoretical models \cite{WengPRX14, FanSciRep17} and experiments including  Angle-Resolved Photoemission Spectroscopy (ARPES)\cite{ManzoniPRL15, MoreschiniPRB16, WuPRX16, ManzoniPRL16, Manzoni17,ZhangNatComm17, XiongPRB17}, Scanning Tunneling Spectroscopy (STS)\cite{WuPRX16, ManzoniPRL16, LiPRL16}, infrared spectroscopy\cite{WangPRB15, WangPRL15, JiangPRB17, WangPNAS17} and quantum oscillations \cite{TianPRB16, XiuNatComm16, TianPRB17, TianPRL17} suggest the system could be a strong topological insulator (TI), weak TI, or Dirac semimetal.  At the same time, transport studies have reported a remarkable set of exotic but seemingly disparate phenomena including a chiral magnetic effect\cite{LiNatPhy16}, anomalous Hall effect\cite{LiangNatPhy18}, discrete scale invariance\cite{WangSciAdv18} three-dimensional quantum Hall effect\cite{TangNat19}, and quantized thermoelectric response\cite{Zhang2020}.  The understanding of the underlying electronic structure that drive these observations is of significant interest.  

It has been pointed out that the electronic structure including the ground state topology are highly sensitive to the lattice constant of \ce{ZrTe5} \cite{WengPRX14, FanSciRep17} which in turn can be affected by the growth method \cite{Chen17,WangSciAdv18}, consistent with experimental results showing the high sensitivity of the system to strain \cite{Chu19}. This calls for comprehensive studies of ZrTe$_5$ single crystals to connect observations of transport exotica with the electronic band structure.  Here, we report an investigation of single crystals grown by chemical vapor transport (CVT) which are characterized by anomalously large and non-monotonic Seebeck and Nernst effects.  We have further performed measurements of electronic transport, magnetic torque, and time and angle resolved photoemission spectroscopy (tr-ARPES) as well as electronic structure calculations.  We show that we can quantitatively describe the thermoelectric, quantum oscillation, and spectroscopic results with a model of a small gap ($\Delta_{gap} = 27 \pm 5$ meV) weak TI with a temperature dependent Fermi level (shifting with rate $k_b\cdot\gamma = 0.48 \pm 0.06$ meV/K).  This establishes a concrete description of ZrTe$_5$ grown in this manner that will further enable band engineering for topology and high thermoelectric performance.

\section{results and discussion}
ZrTe$_5$ crystallizes in the orthorhombic structure of space group $Cmcm$ (D$^{17}_{2h}$, No.63), shown in Fig. \ref{FL_shifting}(a,b). The structure is composed of \ce{ZrTe3} prismatic chains oriented along the $a$-axis with additional Te atoms forming zig-zag between them. A weak inter-chain coupling exists such that two-dimensional layers are formed in the $ac$ plane (denoted by black dashed lines).  The coupling between the layers is van der Waals in nature and is much weaker than the inter-chain coupling; the crystal structure is highly anisotropic in all three directions. The crystals presented here were grown by chemical vapor transport\cite{Chen17} (CVT) and have a morphology characterized by a long ribbon-like shape with a typical dimension of 3 mm $\times$ 0.05 mm $\times$ 0.5 mm along the $a$, $b$, and $c$-axes.

\subsection{Temperature dependence of Fermi level}
\begin{figure} 
\includegraphics[width=0.8\linewidth]{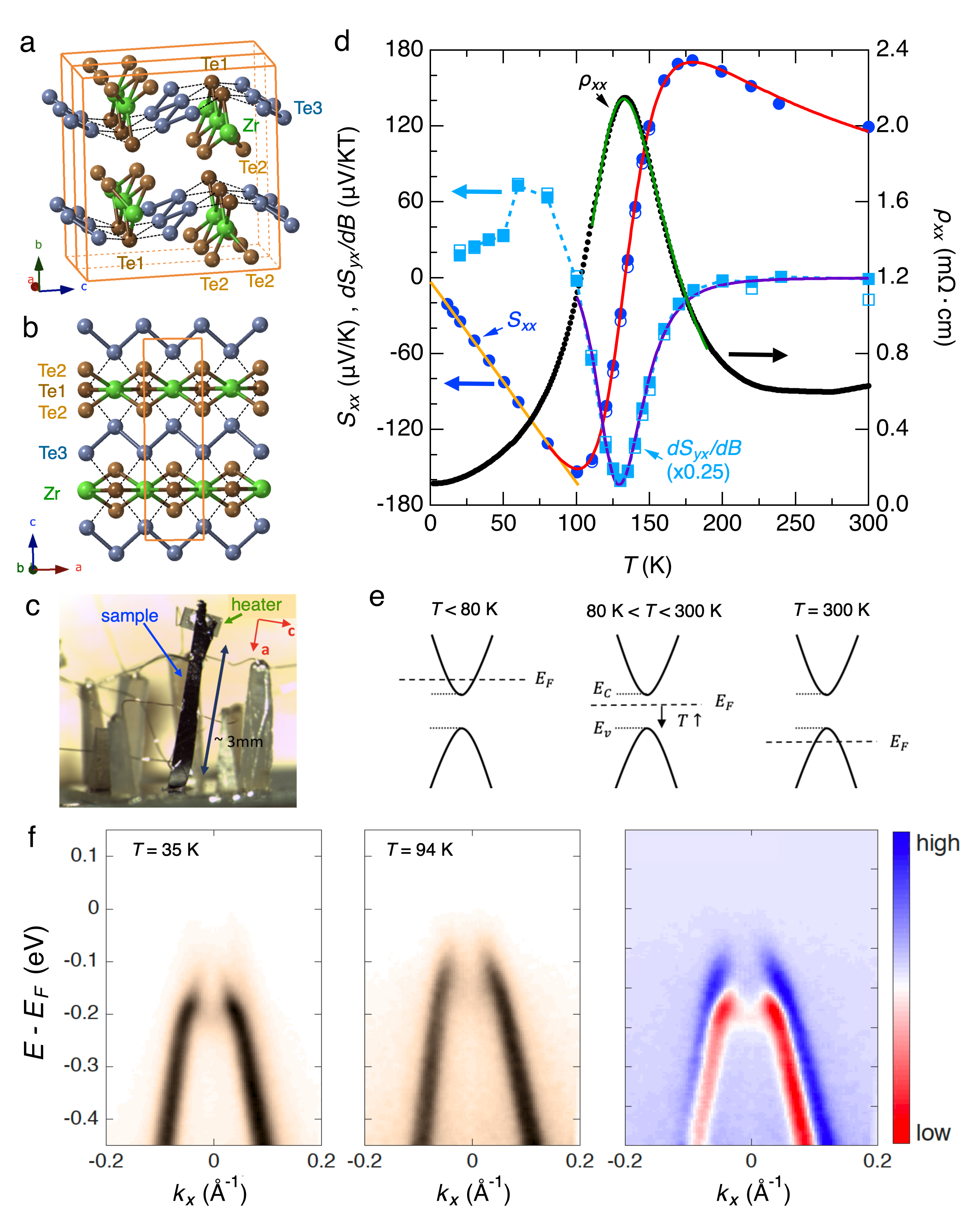}
\caption{\label{FL_shifting}(a,b) Crystal structure of layered compound ZrTe$_5$. (c) Photo of thermoelectric  measurement setup. (d) Resistivity and thermoelectric response for the zero magnetic field limit. The solid and open blue circles (cyan squares) of $S_{xx}$ ($S_{yx}$) refer to data with heater power 0.25 mW and 0.5 mW.  Solid lines in (d) are fittings according to the Fermi level shifting model sketched in (e) (see text).
(f) ARPES Energy-momentum cuts across the $\Gamma$ point taken at temperatures of 35 K (left) and 94 K (middle); darker color denotes higher intensity. The intensity difference (high temperature minus low temperature) is shown in the rightmost panel. }
\end{figure}

We measured the electric and thermoelectric transport response in isothermal and open circuit conditions, respectively. A photo of the thermoelectric measurement setup is shown in Fig. \ref{FL_shifting}(c).  As shown in Fig. \ref{FL_shifting}(d), the zero field longitudinal resistivity $\rho_{xx}$ exhibits a hump structure upon cooling peaking near $T_p=133$ K, as is often reported in CVT grown single crystals\cite{Mcilroy04,Chen17,ZhouPNAS16,TianPRB16,XiuNatComm16,TianPRL17}. 
Concomitant with this feature, the Seebeck coefficient $S_{xx}$ passes through zero. 
The sign change of $S_{xx}$, as well as the low field Hall coefficient \cite{SI} suggest that transport is dominated by hole type carriers for temperature $T > T_p$ and electron type below. 
For temperatures near $T_{p}$ the zero field limit of slope of the Nernst effect $dS_{yx}/dB|_{B=0}$ shows a sign reversal. 
For $T < 80$ K, the system appears to cross over from semiconducting to Fermi liquid behavior suggestive of an electron-like metal.  
More quantitatively, a linear $S_{xx}(T)$ behavior typical for a normal metal is observed, following the Mott formula $S=\frac{\pi^2}{3}\frac{k_B^2T}{e}\left(\frac{\partial\ln\sigma}{\partial \epsilon}\right)_{\epsilon_F}$, where $\epsilon_F$ is the Fermi level with respect to the band bottom and $\sigma$ is electron conductivity.
A linear fit yields $\frac{dS_{xx}}{dT}=-1.59$ $\mu$V/K$^2$.  In addition, the electrical response is well described by parabolic law\cite{SI} $\rho_{xx}=\rho_0+A\cdot T^{2}$, where $A=0.05\pm0.02$ $\mu\Omega\cdot {\rm cm}/{\rm K}^{2}$.  
Qualitatively, this can be explained by a simple semiconducting band structure and $T$-dependent Fermi level as sketched Fig. \ref{FL_shifting}(e).

We have performed angle-resolved photoemission spectroscopy (ARPES) on crystals from this same batch.  Shown in Fig. \ref{FL_shifting}(f) are energy-momentum cuts across the $\Gamma$ point taken at $T = $ 35 K and $T = $ 94 K, as well as their difference.  While the band shape remains largely unchanged, a clear downwards energy shifting at lower $T$ is observed.  Such a shift has been observed in a number of recent reports on CVT grown crystals\cite{ZhangNatComm17,XiongPRB17}.  This is qualitatively consistent with the rigid band shift depicted in  Fig. \ref{FL_shifting}(e).


Significant attention has been aimed at understanding the nature of the band gap in \ce{ZrTe5} including if it is gapped or gapless \cite{ZhangNatComm17,XiongPRB17}.  We show  here that the experimental observations above can be quantitatively captured by a gapped scenario depicted in Fig. \ref{FL_shifting}(e).  Starting with the assumption of a symmetric semiconducting band structure with a gap $\Delta_{gap}$ and Fermi level at $\epsilon_F$, from the Drude model the low field Seebeck and Nernst coefficients are
\begin{equation}
S_{xx} = \frac{k_B}{e}\cdot\frac{pA^h-nA^e}{p+n},~~
S_{yx} = \frac{k_B}{e}\cdot\frac{-2pn\left(A^h+A^e\right)}{p+n}\cdot\omega_c\tau_m
\end{equation}
where $k_B$ is Boltzmann constant, $e$ is free electron charge, $\omega_c$ is the cyclotron frequency, $\tau_m$ is the momentum relaxation time, and $\frac{k_B}{e}A^{e(h)}=\frac{k_B}{e}\left(\frac{|\Delta \epsilon|}{k_BT}+1\right)$ corresponds to the single carrier type Seebeck coefficient.  Given the semiconducting structure, the electron (hole) density $n$ ($p$) are proportional to $\exp{\frac{-|\Delta \epsilon|}{k_BT}}$, with  $|\Delta \epsilon|$ the energy difference between band top (bottom) and the Fermi level.


Previous ARPES observations suggest that the Fermi level shifting is approximately linear in energy in the intermediate $T$ range of the resistivity anomaly \cite{ZhangNatComm17}. We approximate $\epsilon_F-\epsilon_m \approx -\gamma k_B (T - T_m)$, where $\epsilon_m$ denotes the mid point of band gap, $T_m$ denotes the temperature when Fermi level is degenerate with the midpoint,  and $\gamma$ is defined as a dimensionless shifting rate.  Applying this to the model above, we arrive at expressions for the low field thermoelectric coefficients
\begin{eqnarray}
S_{xx} &=& \frac{k_B}{e}\cdot\left(\frac{\Delta_{gap}}{2k_BT}+\frac{5}{2}-\beta\right)\cdot\tanh\left(\gamma\frac{T-T_m}{T}\right)-\frac{k_B}{e}\cdot\frac{\gamma(T- T_m)}{T},\\
S_{yx} &=& \frac{-k_B}{e}\cdot\left(\frac{\Delta_{gap}}{2k_BT}+\frac{5}{2}-\beta\right)\cdot\cosh^{-2}\left(\gamma\frac{T-T_m}{T}\right)\cdot\omega_c\tau_m
\end{eqnarray}

Whereas a single-band semiconductor with a static Fermi level the thermoelectric coefficients will be monotonic and retain the same sign at different $T$, the shifting of $\epsilon_{F}$ allows for a dynamic response. The corresponding electrical response \cite{SI} is
\begin{equation}
\rho_{xx}=\frac{m^*}{(p+n)e^2\tau_m}\propto \frac{m^*}{e^2\tau_m}\cdot \exp\left(\frac{\Delta_{gap}}{2k_BT}\right)\cdot\cosh^{-1}\left(\gamma\frac{T-T_m}{T}\right)
\end{equation}
\begin{equation}
\rho_{yx}=\frac{B(p-n)}{e(p+n)^2}\propto\frac{B}{e}\cdot \exp\left(\frac{\Delta_{gap}}{2k_BT}\right)\cdot\tanh\left(\gamma\frac{T-T_m}{T}\right)\cdot\cosh^{-1}\left(\gamma\frac{T-T_m}{T}\right)
\end{equation}
Similar to the case of the thermoelectric response, the additional degree of freedom associated with shifting $\epsilon_{F}$ makes an important modification to the electrical transport response as a function of $T$.



We compare the expectations for Eq. (2)-(5) to the experimental results in Fig. \ref{FL_shifting}(d).
We directly fit $\rho_{xx}(T)$ and $S_{xx}(T)$ and for the Nernst and Hall response we fit the low field slope $\left.\frac{dS_{yx}}{dB} \right|_{B \to 0}$ and $\left. \frac{d\rho_{yx}}{dB} \right|_{B \to 0}$.
The fit captures the intermediate and high $T$ response with fit parameters and error bars listed in Table \ref{tableFit}.  The obtained $k_b\cdot\gamma = 0.48\pm0.06$ meV/K is consistent with recent ARPES reports\cite{ZhangNatComm17}, where the band shifting rate is approximately $0.43$ meV/K . The obtained $\Delta_{gap}= 27\pm5$ meV is also consistent with our time-resolved ARPES, which we discuss below. The error bar above is defined as the standard deviation among the results in Table \ref{tableFit}.


\begin{table}[htp]
\caption{\label{tableFit}Comparison of fitting parameters for the Fermi level shifting rate $\gamma$, zero-field energy gap $\Delta_{\textrm{gap}}$, mid-point temperature $T_{m}$, and power law of inverse scattering time $\tau$ for three samples s0, s1, and s2. Fitting to the zero field resistivity $\rho_{xx}$, low field slope of the Nernst response $dS_{yx}/dB$, zero field Seebeck coefficient $S_{xx}$, and low field Hall slope $d\rho_{yx}/dB$ are shown.} 
\begin{center}
\resizebox{\columnwidth}{!}{%
\renewcommand{\arraystretch}{0.5}
\begin{tabular}{|c|c|c|c|c|c|}
\hline
  sample & s0 & s1 & s1 & s2 & s2 \\
\hline
  Fitting & $ \rho_{xx}(B=0)$ & $(dS_{yx}/dB)_{B\rightarrow0}$ & $S_{xx}(B=0)$ & $ \rho_{xx}(B=0)$ & $(d\rho_{yx}/dB)_{B\rightarrow0}$\\
  \hline
  Fit regime & 110-190 K  & 100-300 K & 80-300 K & 110-190 K & 100-180 K \\
  \hline
  $k_b\cdot\gamma~(\textrm{meV/K})$ & $0.42\pm0.01$ & $0.55\pm0.02$ & $0.53\pm0.02$& $0.417\pm0.004$& $0.46\pm0.03$\\
 $\Delta_{\textrm{gap}}~(\textrm{meV})$ & $25\pm27$ &  -  &  $24\pm2$& $ 33\pm8 $ &- \\
$T_m~(\textrm{K})$ & $136.6\pm0.9$ & $ 130\pm2 $ & $ 137.0\pm0.4 $ & $142.1\pm0.2 $& $127\pm1 $\\
$\beta~(\tau\propto T^{-\beta})$ & $1.8\pm1.0$ & $0.2\pm1.1$ & $1.00\pm0.08$& $1.7\pm0.3$&-\\
\hline
\end{tabular}
}
\end{center}
\end{table}

For $T<80$ K, this semiconducting model fails to capture the transport response, as would be expected upon entering the metallic regime.  Moreover, as noted above $\rho_{xx}(T)$ and $S_{xx}(T)$ are captured by that expected from a simple metal, suggesting a drop in the magnitude of $\gamma$.  
We hypothesize that this may be due to the increase in the density of states in the conduction band at the $\Gamma$ point, or that the shifting itself is driven by lattice contraction which would be expected to saturate at low $T$ when the thermal expansion coefficient vanishes \cite{Kittel}.  




\subsection{Fermiology of conduction band}

\begin{figure} 
\includegraphics[width=0.8\linewidth]{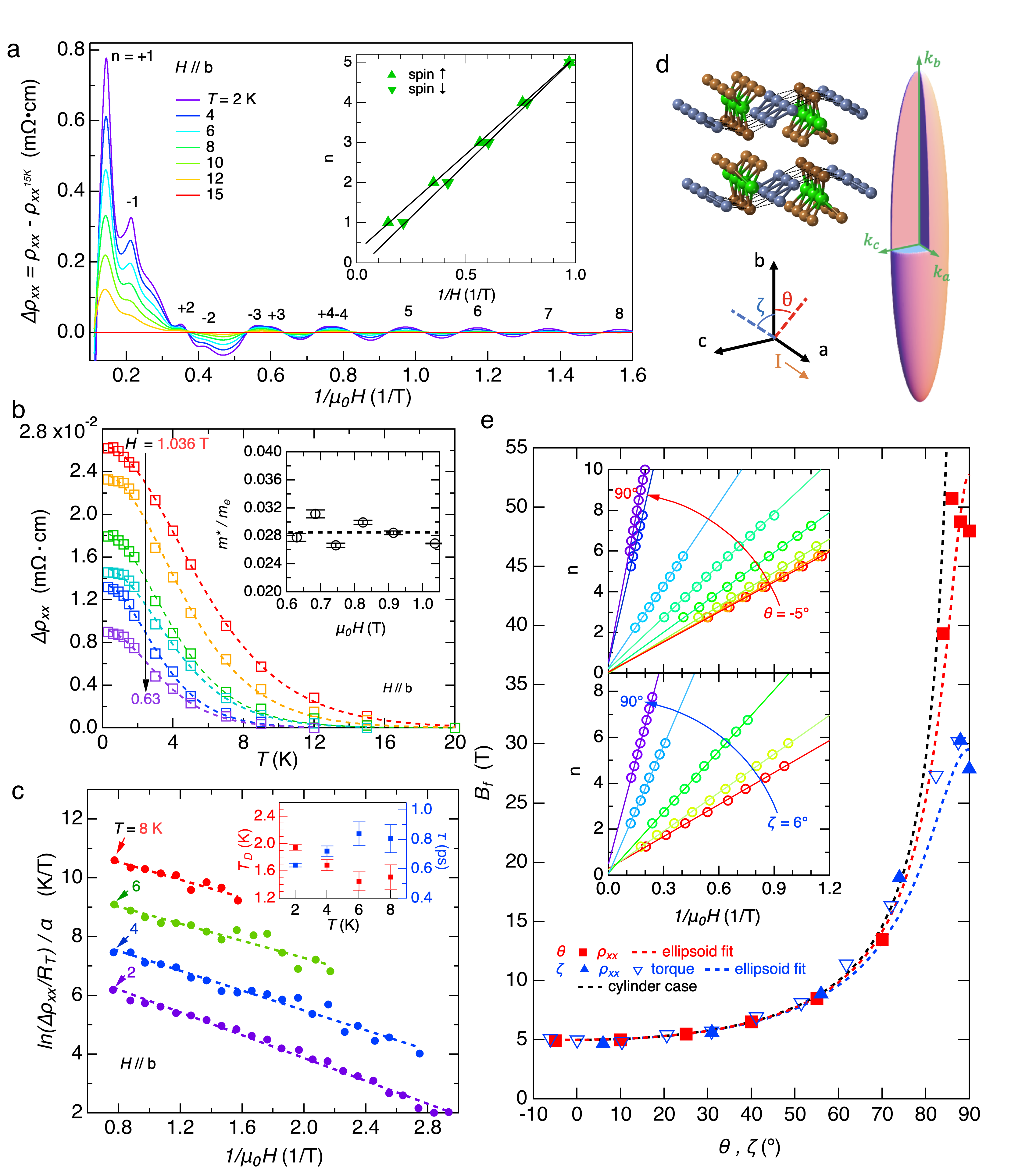}%
\caption{\label{SdH}(a) Oscillatory magnetoresistance $\Delta \rho_{xx} \equiv \rho_{xx}(H,T) - \rho_{xx}(H,T=15$ K$)$.  The inset shows the spin-split Landau fan diagram.
(b) Oscillation amplitude $\Delta\rho_{xx}$(T) at fixed field with $H\|b$. Dashed lines are fittings according to thermal damping (see text) and the fitted effective mass is plotted in the inset.
(c) Dingle analysis fits, where $\alpha=\frac{2\pi^2k_Bm^*}{\hbar e}$. The fitted Dingle temperature $T_D$ and carrier lifetime $\tau$ are plotted in the inset.
(d) Illustration of measurement configuration for quantum oscillations and relation to crystal structure along with a schematic depiction of the observed ellipsoidal Fermi surface.
(e) Quantum oscillation frequency as a function of field angle. Red squares (blue triangles) refer to rotating from $b$ to $a$ ($c$) axis. Both transport (solid) and torque (open) results are plotted. Dashed lines are fits to either ellipsoid or cylinder Fermi surface. Landau fan diagrams are shown in the insets.
}
\end{figure}

We further verify the above description of CVT grown \ce{ZrTe5} by studying the Fermiology of the low temperaure band structure with quantum oscillations and tr-ARPES.  For the former we examine the low temperature magnetoresistance $\Delta \rho_{xx}(H,T) \equiv \rho_{xx}(H,T) - \rho_{xx}(H,T=15$ K$)$, \textit{i.e.} using the $T=15$ K trace as a background.  
This is shown plotted against $1/H$ in Fig.  \ref{SdH}(a) for field applied along the $b$ axis.  Clear Shubnikov-de Haas (SdH) oscillations are observed, onsetting near $\mu H_0\sim0.25$ T, indicating a quantum  mobility exceeding $\mu=1/\mu_{0} H_0\sim4\times10^4$ cm$^2$/Vs.
The oscillation frequency $B_{f}^{\hat{i}}$ is $5.070\pm0.005$ T which corresponds to a Fermi surface cross section $A_F^{ac}=4.84\pm 0.01\times10^{-4}$ \AA$^{-2}$. 

We have measured the detailed angular dependence of the SdH effect as well as de Haas-van Alphen oscillations in magnetic torque \cite{SI}; the variation of oscillation frequency $B_{f}^{\hat{i}}$ as the field rotates away from $b$ axis are plotted in Fig. \ref{SdH}(e), where red squares (blue triangles) denotes rotating in $ab$ ($bc$ plane) and $\theta$ ($\zeta$) denotes the angle between $b$ axis and field direction. The angular variation is consistent with an ellipsoidal (3D) rather than cylindrical (quasi-2D) Fermi surface (dashed lines).  This is also emphasized by the corresponding Landau fan diagrams (Fig. \ref{SdH}(e) inset) in which a saturating slope for $H\| b$ is observed.  Assuming an ellipsoid Fermi surface, the corresponding  Fermi wave vectors are $k_F^a=9.46\times10^{-3}~$\AA$^{-1}$, $k_F^c=1.63\times10^{-2}~$\AA$^{-1}$ and $k_F^b=9.32\times10^{-2}~$\AA$^{-1}$ (depicted in Fig. \ref{SdH}(d)). 
This ellipsoid Fermi pocket corresponds to a carrier density of $n=\frac{2}{(2\pi)^3}V_{FS}=4.85\times10^{17}$ cm$^{-3}$, in good agreement wiht that obtained from the Hall coefficient  $n=4.71\times10^{17}$ cm$^{-3}$~ \cite{SI}.  Together, these provide a consistent picture of the low $T$ Fermi surface being composed of a single electron pocket.

We can provide a further quantitative comparison with analysis of cyclotron effective mass $m^*$, carrier lifetime $\tau$, and effective $g$ factor from the Lifshitz-Kosevich formula\cite{Shoenberg}:
\begin{equation}
\Delta \rho_{xx} \propto R_TR_DR_S\cdot \cos\left(2\pi(\frac{B_{f}^{\hat{i}}}{B}+\phi)\right)
\end{equation}
The oscillation amplitude is modulated by three factors: the thermal factor $R_T=\frac{\alpha T}{B}/\sinh(\frac{\alpha T}{B})$ due to thermal broadening, the Dingle factor $R_D=\exp\left(\frac{-\alpha T_D}{B}\right)$ due to scattering, and the spin factor $R_S=\cos\left(\frac{\pi gm^*}{2m_e}\right)$ due to Zeeman splitting, where $\alpha=\frac{2\pi^2k_Bm^*}{e\hbar}$.
The oscillation amplitude $\Delta\rho_{xx}(T)$ at fixed field ($H\|b$) is plotted in Fig. \ref{SdH}(b), with a fit to $R_T$ yielding an average $ac$-plane $m^*_{ac}=0.028m_e$ (the result of each field is shown inset) and  Fermi velocity $v_F^{ac}=\hbar \langle k_F^{ac}\rangle/m_{ac}^*=4.1\times10^5$ m/s.  As shown in Fig.  \ref{SdH}(c), an average Dingle temperature $T_D=1.65$ K is found from a linear fit to $\ln\left(\Delta\rho_{xx}/R_T\right)/\alpha\propto \ln(R_D) /\alpha = \frac{- T_D}{B} $.  From $T_D=\frac{\hbar}{2\pi k_B \tau}$ this corresponds to a lifetime $\tau=0.75$ ps, comparable to $\tau_m=1.17$ ps calculated from the Hall mobility and $m^*_{ac}$.

We note that the analysis of above is restricted to $\mu_{0} H < 1.3$ T, as in larger field a pronounced Zeeman splitting becomes evident, as shown in Fig. \ref{SdH}(a) and its inset. The modulation effect due $R_S$ can be rewritten as 
\begin{equation}
R_S\cos\left(2\pi(\frac{B_f}{B}+\phi)\right)=\frac{1}{2}\sum_{\uparrow,\downarrow} \cos\left(2\pi(\frac{B^{\uparrow(\downarrow)}_f}{B}+\phi\pm\frac{gm^*}{4m_e})\right)
\end{equation}
where the summation is over two spins (+(-) for spin up $\uparrow$ (down $\downarrow$)). 
Therefore, the spin split Landau level indeces correspond to the  lines with $n_{\uparrow(\downarrow)}=\frac{B^{\uparrow(\downarrow)}_f}{B}+\phi \pm \frac{gm^*}{4m_e}$, as shown in the inset of Fig. \ref{SdH}(a) for $H\|b$. 
Linear fitting gives $B^\uparrow_f=4.85\pm0.04$ T (intercept=$0.29\pm0.02$) and $B^\downarrow_f=5.31\pm0.07$ T (intercept=$-0.16\pm0.04$), corresponding to $g \approx 25.8$. 

\begin{figure}
\includegraphics[width=0.8\linewidth]{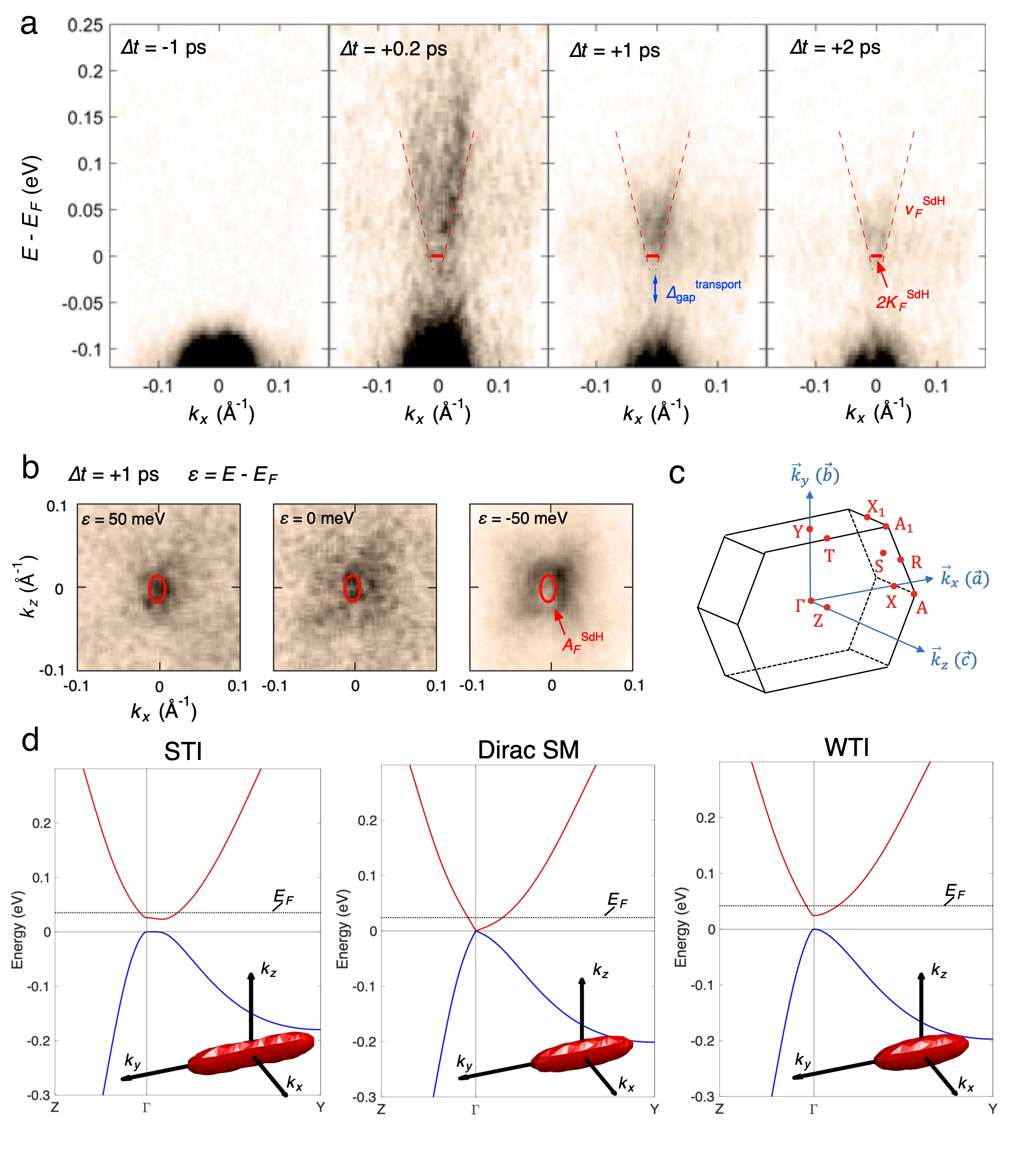}%
\caption{\label{ARPES} (a) tr-ARPES energy-momentum cuts across the $\Gamma$ point at different time delays, probed at $T = 35$ K.  $k_x$ is along the $a$-axis and $k_x=0$ corresponds to the $\Gamma$ point. At the Fermi level, the solid red bar and dashed line refer to $k^{a}_F$ and $v_F^{ac}$ determined from quantum oscillations, respectively. The gap size $\Delta_{gap}$ is determined from the transport data analysis is also drawn. 
(b) Energy contours obtained 1 ps after photoexcitation, at energy $\epsilon= E-E_F=50$, 0, -50 meV, where $k_x$ is along the $a$ axis and $k_z$ is along the $c$ axis. The integration window is 50 meV. The red ellipse represents the Fermi surface size obtained from quantum oscillations.
(c) Brillioun zone and (d) band structure calculated by density functional theory. The left, middle and right panels correspond to strong topological insulator (TI), Dirac semimetal (SM) and weak TI phases, respectively. For each scenario, the low temperature Fermi level is denoted by the dashed line, determined by the Fermi surface volume from quantum oscillations. The corresponding Fermi pocket is depicted in the inset.}
\end{figure}

Turning to spectroscopy of the conduction band, results of tr-ARPES performed at $T = 35$ K are shown in Fig. \ref{ARPES}(a,b).  
Here, the conduction band at $\epsilon_{F}$ and the band gap $\Delta_{gap}$ are clearly resolved.   The horizontal red bar and dashed red lines in Fig. \ref{ARPES}(a) indicate the magnitude of  Fermi wave vector $k^{a}_F$ and Fermi velocity $v_F^{ac}$ obtained from quantum oscillations, which are in approximate agreement. $\Delta_{gap}$ derived from the transport data in Fig. \ref{FL_shifting}(d) is also drawn in Fig. \ref{ARPES}(a).  In Fig. \ref{ARPES}(b) the Fermi surface cross section is again compared with an in-plane spectrum, showing good agreement at $\epsilon_{F}$.  

Using the Brillion zone defined in Fig. \ref{ARPES}(c), in Fig. \ref{ARPES}(d) we show the band structure at $\Gamma$ point of the three possible phases for \ce{ZrTe5}, namely strong TI (left panel), Dirac semimetal (middle), and weak TI (right), calculated by density functional theory (DFT). The electronic topology of \ce{ZrTe5} is known to be extremely sensitive to the values of its lattice constants\cite{FanSciRep17}; here, the three panels in Fig. \ref{ARPES}(d) are obtained in series thru expanding unit cell volume by $8\times10^{-7}\%$, from left to right (the direct gap for the strong TI and weak TI phase nominally agree with experimental observations). 
In each, the Fermi level is shown for which the Fermi surface volume matches that of the quantum oscillation analysis (dashed line). The corresponding Fermi pockets are plotted in the insets. 
The band inversion of the  strong TI leads to a weaker dispersion; comparing the flatness of these ellipsoids, the strong TI case is closer to that from quantum oscillation results depicted in Fig.\ref{SdH}(d).  However,  surface states were not seen in the tr-ARPES measurements (nor \textit{e.g.} previous static ARPES reports with high energy resolution\cite{ZhangNatComm17,XiongPRB17}). 
One possible scenario for this within a strong TI system is leakage of surface state into the bulk due to small band gap. The expected inelastic mean free path of excited electrons here is 2-7 nm compared to the penetration depth of surface states estimated to be 45 nm from $\Delta_{gap}=\frac{(\hbar k)^2}{2m^*}$.  Additional experiments to probe the potential surface states are of significant interest.

\subsection{Field Enhancement of $S_{xx}$ and $S_{yx}$}
\begin{figure} 
\includegraphics[width=0.8\linewidth]{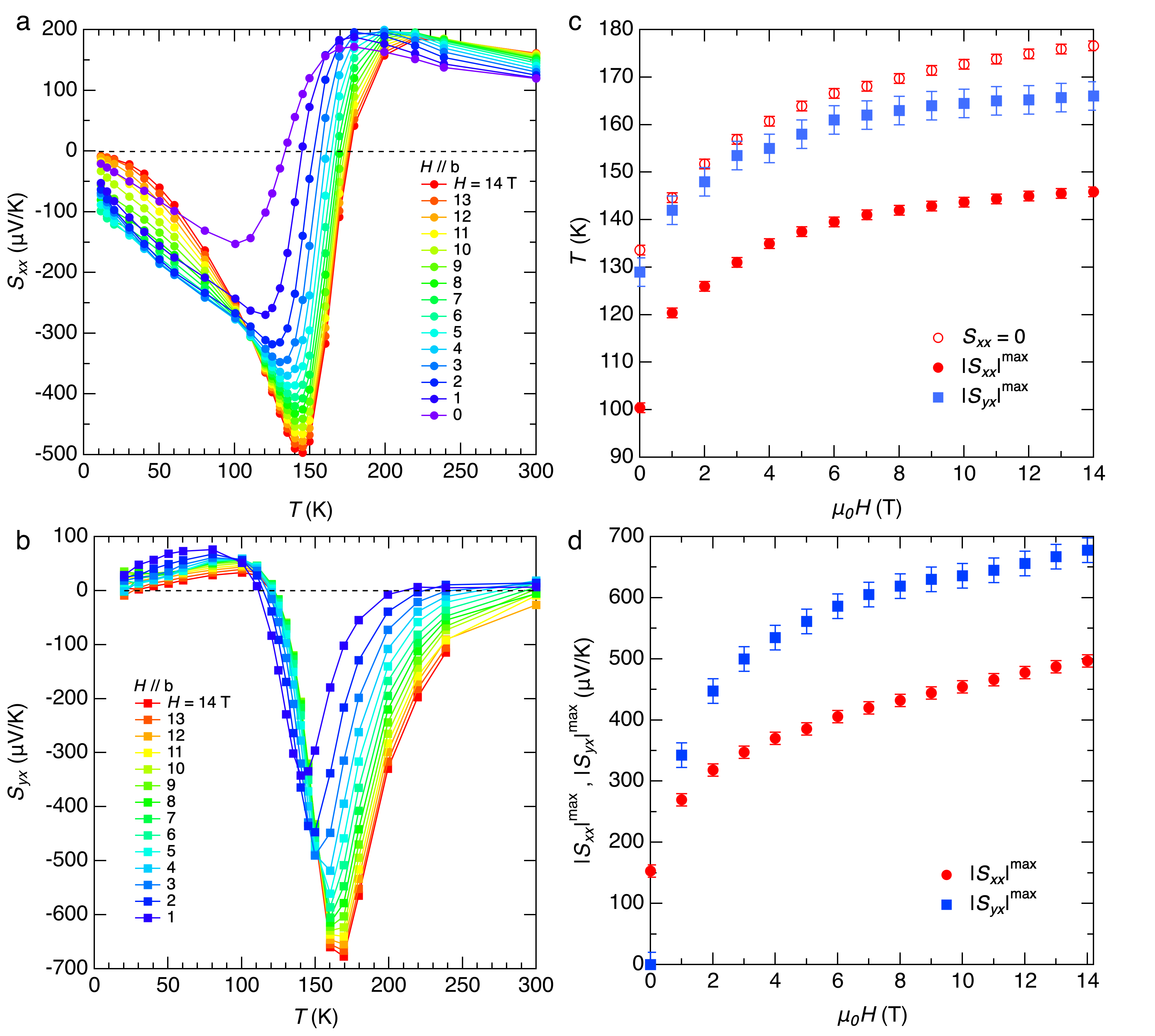}
\caption{\label{Sxx_H}(a) Thermopower $S_{xx}$ and (b) Nernst coefficient $S_{yx}$ as a function of temperature in different magnetic field $H$. (c)  Shifting of the $S_{xx}$ and $S_{yx}$ peak temperature with $H$ as well as the $H$-dependence of the zero-crossing of $S_{xx}(T)$. (d) The maximum value of $S_{xx}$ and $S_{yx}$ as a function of magnetic field. The error bars in (c) and (d) reflect the discreteness of the field-dependent data set.}
\end{figure}

At 100 K $< T <$ 150 K where the thermoelectric response changes most rapidly, we observe a strong enhancement of both $S_{xx}$ and $S_{yx}$ with increasing $H\|b$.  
As shown in  Fig. \ref{Sxx_H}(a,b), peak magnitude of $S_{xx}(T)$ is enhanced more than 3 fold at 14 T, reaching 500 $\mu$V/K, while that of $S_{yx}$ approaches  700 $\mu$V/K. The maximum values at fixed $H$ for both quantities appear to be monotonically increasing with $H$ without saturation at our largest applied fields (see Fig. \ref{Sxx_H}(d)).  
The Seebeck coefficient of a simple metal is not typically strongly enhanced in field; this suggests a significant modification to the electronic band structure.  Given the acute sensitivity of $\epsilon_F$ to $T$, one natural origin for this response would be a magnetic field dependence to $\epsilon_F$.  To connect these, we plot the $H$ dependence of the $T$ at which the thermoelectric coefficients peaked and the zero crossing for $S_{xx}(T)$ occur.  
They show similar behavior with features shifting towards higher temperature for stronger magnetic fields, consistent with a spin down valence band being raised by the Zeeman energy at high temperature. Upon cooling, this would lead to the Fermi level traversing the gap at higher $T$, which can be tracked by a vanishing  $S_{xx}(T)$ and maximal $S_{yx}(T)$. From zero field to $H=14$ T, the observed shift is approximately 40 K, which corresponds to a Fermi level shift of $\approx 20$ meV given the fitted shift rate $\gamma$ from Table \ref{tableFit}. This is comparable to the size of the Zeeman splitting $\approx 20$ meV given the large $g$ factor.  
We note also this is approaching the size of the zero-field energy gap, suggesting a potentially complex evolution at high field.  Thus despite their unusually large magnitude, these effects are consistent with the band structure described above.  It is of significant interest to pursue these and related materials to higher magnetic fields to test recent predictions for extremely large field-induced figure of merit $ZT$ in systems with highly dispersive bands\cite{Liang18}.

\section{Conclusion}
We have synthesized single crystals of ZrTe$_5$ by chemical vapor transport and probed the band properties via electric and thermoelectric transport, magnetic torque, and tr-ARPES. Both transport and photoemission results indicate a temperature dependent Fermi level shifting across a small semiconducting gap. A semiclassical calculation of the transport coefficients based on such a model describes the observations, yielding $k_b\cdot\gamma= 0.48\pm0.06$ mev/K and $\Delta_{gap}=27\pm5$ meV.  For $T$ below this range, the system behaves as a simple metal.  Quantum oscillations and ARPES reveal a consistent picture of an a light-mass (0.028$m_e$), small ellipsoidal electron pocket with large  $v_{F} \approx 4\times10^5$ m/s.  The lack of surface states in ARPES is consistent with weak TI regime, supported by our first principles calculations.  Finally, a significant enhancement of both the Seebeck and Nernst effect with magnetic field indicates that the band structure is strongly affected in $H$, consistent with the relatively large $g = 25.8$.  Together this study demonstrates that comprehensive experiments and analysis of single crystals of \ce{ZrTe5} grown in the same manner can produce a clear picture of the underlying physical mechanisms for unusual transport effects.  Extending these studies to the recent exotica reported \cite{LiNatPhy16,LiangNatPhy18,WangSciAdv18,TangNat19,Zhang2020} in \ce{ZrTe5} grown by other methods may elucidate the origin of the rich electronic physics in this system.


\section{Methods}

\subsection{Crystal Growth}
Polycrystalline \ce{ZrTe5} was obtained by sintering high purity zirconium (99.9999\%) and tellurium (99.9999\%) at 500 \degree C .
Single crystals were grown by a CVT method with the prepared polycrystalline powder and \ce{I2} as the transport agent. The temperature of source zone and growth zone are 520 \degree C and 475 \degree C, respectively. 
The obtained crystal were confirmed as structurally single phase by powder X-ray diffraction.

\subsection{Measurement}
Electrical and thermoelectric transport measurements were performed in commercial cryostats equipped with superconducting magnets. Electrical transport is measured with a standard Hall bar configuration of six gold wire contacts. Thermoelectric transport is measured with a static temperature gradient along the $a$ axis driven a heater directly attached to one end of the crystal. Temperature gradient is detected by a pair of thermocouple and electrical voltage drops are detected by two pairs of electrical contacts along $a$ and $c$ axis which gives Seebeck and Nernst signal respectively. Torque measurements are done with a home-made cantilever magnetometer. 

Photoemission measurement was done at 6 eV photon energy. The sample was cleaved \textit{in situ} under a pressure of $1\times10^{-10}$ torr and a temperature of 35 K. In time-resolved measurements the system was optically pumped with 1.6 eV laser pulses before subsequent probe pulses.

\subsection{Density Functional Theory Calculations}
We performed the density functional theory (DFT) calculations implemented in Vienna Ab initio Simulation Package (VASP) code ~\cite{vasp1,vasp2}, with the Projector augmented wave method (PAW) \cite{PAW} for the pseudo potential formalism. The ground state for the ZrTe$_5$ primitive unit cell is converged with exchange-correlation energies parametrized by Perdew, Burke and Ernzerhof (PBE)~\cite{pbe}, a 300 eV cutoff energy for the plane-wave-basis set, and a $15 \times 15 \times 7$ Monkhorst-Pack grid sampling~\cite{MP_grid} in the reciprocal space. The $Z_2$ topological invariants are computed using the Fu-Kane parity formula with inversion symmetry~\cite{FuKane_parity}.



\section{Data Availability}
The data that support the findings of this study are available from the corresponding author on reasonable request.

\bibliography{ZrTe5CVT_v2}

\begin{thebibliography}{40}%
\makeatletter
\providecommand \@ifxundefined [1]{%
 \@ifx{#1\undefined}
}%
\providecommand \@ifnum [1]{%
 \ifnum #1\expandafter \@firstoftwo
 \else \expandafter \@secondoftwo
 \fi
}%
\providecommand \@ifx [1]{%
 \ifx #1\expandafter \@firstoftwo
 \else \expandafter \@secondoftwo
 \fi
}%
\providecommand \natexlab [1]{#1}%
\providecommand \enquote  [1]{``#1''}%
\providecommand \bibnamefont  [1]{#1}%
\providecommand \bibfnamefont [1]{#1}%
\providecommand \citenamefont [1]{#1}%
\providecommand \href@noop [0]{\@secondoftwo}%
\providecommand \href [0]{\begingroup \@sanitize@url \@href}%
\providecommand \@href[1]{\@@startlink{#1}\@@href}%
\providecommand \@@href[1]{\endgroup#1\@@endlink}%
\providecommand \@sanitize@url [0]{\catcode `\\12\catcode `\$12\catcode
  `\&12\catcode `\#12\catcode `\^12\catcode `\_12\catcode `\%12\relax}%
\providecommand \@@startlink[1]{}%
\providecommand \@@endlink[0]{}%
\providecommand \url  [0]{\begingroup\@sanitize@url \@url }%
\providecommand \@url [1]{\endgroup\@href {#1}{\urlprefix }}%
\providecommand \urlprefix  [0]{URL }%
\providecommand \Eprint [0]{\href }%
\providecommand \doibase [0]{http://dx.doi.org/}%
\providecommand \selectlanguage [0]{\@gobble}%
\providecommand \bibinfo  [0]{\@secondoftwo}%
\providecommand \bibfield  [0]{\@secondoftwo}%
\providecommand \translation [1]{[#1]}%
\providecommand \BibitemOpen [0]{}%
\providecommand \bibitemStop [0]{}%
\providecommand \bibitemNoStop [0]{.\EOS\space}%
\providecommand \EOS [0]{\spacefactor3000\relax}%
\providecommand \BibitemShut  [1]{\csname bibitem#1\endcsname}%
\let\auto@bib@innerbib\@empty
\bibitem [{\citenamefont {Weng}\ \emph {et~al.}(2014)\citenamefont {Weng},
  \citenamefont {Dai},\ and\ \citenamefont {Fang}}]{WengPRX14}%
  \BibitemOpen
  \bibfield  {author} {\bibinfo {author} {\bibfnamefont {H.}~\bibnamefont
  {Weng}}, \bibinfo {author} {\bibfnamefont {X.}~\bibnamefont {Dai}}, \ and\
  \bibinfo {author} {\bibfnamefont {Z.}~\bibnamefont {Fang}},\ }\href {\doibase
  10.1103/PhysRevX.4.011002} {\bibfield  {journal} {\bibinfo  {journal} {Phys.
  Rev. X}\ }\textbf {\bibinfo {volume} {4}},\ \bibinfo {pages} {011002}
  (\bibinfo {year} {2014})}\BibitemShut {NoStop}%
\bibitem [{\citenamefont {Okada}\ \emph {et~al.}(1980)\citenamefont {Okada},
  \citenamefont {Sambongi},\ and\ \citenamefont {Ido}}]{okada1980giant}%
  \BibitemOpen
  \bibfield  {author} {\bibinfo {author} {\bibfnamefont {S.}~\bibnamefont
  {Okada}}, \bibinfo {author} {\bibfnamefont {T.}~\bibnamefont {Sambongi}}, \
  and\ \bibinfo {author} {\bibfnamefont {M.}~\bibnamefont {Ido}},\ }\href@noop
  {} {\bibfield  {journal} {\bibinfo  {journal} {Journal of the Physical
  Society of Japan}\ }\textbf {\bibinfo {volume} {49}},\ \bibinfo {pages} {839}
  (\bibinfo {year} {1980})}\BibitemShut {NoStop}%
\bibitem [{\citenamefont {Okada}\ \emph {et~al.}(1982)\citenamefont {Okada},
  \citenamefont {Sambongi}, \citenamefont {Ido}, \citenamefont {Tazuke},
  \citenamefont {Aoki},\ and\ \citenamefont {Fujita}}]{Okada1982}%
  \BibitemOpen
  \bibfield  {author} {\bibinfo {author} {\bibfnamefont {S.}~\bibnamefont
  {Okada}}, \bibinfo {author} {\bibfnamefont {T.}~\bibnamefont {Sambongi}},
  \bibinfo {author} {\bibfnamefont {M.}~\bibnamefont {Ido}}, \bibinfo {author}
  {\bibfnamefont {Y.}~\bibnamefont {Tazuke}}, \bibinfo {author} {\bibfnamefont
  {R.}~\bibnamefont {Aoki}}, \ and\ \bibinfo {author} {\bibfnamefont
  {O.}~\bibnamefont {Fujita}},\ }\href {\doibase 10.1143/JPSJ.51.460}
  {\bibfield  {journal} {\bibinfo  {journal} {Journal of the Physical Society
  of Japan}\ }\textbf {\bibinfo {volume} {51}},\ \bibinfo {pages} {460}
  (\bibinfo {year} {1982})},\ \Eprint
  {http://arxiv.org/abs/https://doi.org/10.1143/JPSJ.51.460}
  {https://doi.org/10.1143/JPSJ.51.460} \BibitemShut {NoStop}%
\bibitem [{\citenamefont {Jones}\ \emph {et~al.}(1982)\citenamefont {Jones},
  \citenamefont {Fuller}, \citenamefont {Wieting},\ and\ \citenamefont
  {Levy}}]{Jones1982Sxx}%
  \BibitemOpen
  \bibfield  {author} {\bibinfo {author} {\bibfnamefont {T.}~\bibnamefont
  {Jones}}, \bibinfo {author} {\bibfnamefont {W.}~\bibnamefont {Fuller}},
  \bibinfo {author} {\bibfnamefont {T.}~\bibnamefont {Wieting}}, \ and\
  \bibinfo {author} {\bibfnamefont {F.}~\bibnamefont {Levy}},\ }\href@noop {}
  {\bibfield  {journal} {\bibinfo  {journal} {Solid State Communications}\
  }\textbf {\bibinfo {volume} {42}},\ \bibinfo {pages} {793} (\bibinfo {year}
  {1982})}\BibitemShut {NoStop}%
\bibitem [{\citenamefont {Fan}\ \emph {et~al.}(2017)\citenamefont {Fan},
  \citenamefont {Liang}, \citenamefont {Chen}, \citenamefont {Yao},\ and\
  \citenamefont {Zhou}}]{FanSciRep17}%
  \BibitemOpen
  \bibfield  {author} {\bibinfo {author} {\bibfnamefont {Z.}~\bibnamefont
  {Fan}}, \bibinfo {author} {\bibfnamefont {Q.-F.}\ \bibnamefont {Liang}},
  \bibinfo {author} {\bibfnamefont {Y.}~\bibnamefont {Chen}}, \bibinfo {author}
  {\bibfnamefont {S.-H.}\ \bibnamefont {Yao}}, \ and\ \bibinfo {author}
  {\bibfnamefont {J.}~\bibnamefont {Zhou}},\ }\href@noop {} {\bibfield
  {journal} {\bibinfo  {journal} {Scientific reports}\ }\textbf {\bibinfo
  {volume} {7}},\ \bibinfo {pages} {45667} (\bibinfo {year}
  {2017})}\BibitemShut {NoStop}%
\bibitem [{\citenamefont {Manzoni}\ \emph {et~al.}(2015)\citenamefont
  {Manzoni}, \citenamefont {Sterzi}, \citenamefont {Crepaldi}, \citenamefont
  {Diego}, \citenamefont {Cilento}, \citenamefont {Zacchigna}, \citenamefont
  {Bugnon}, \citenamefont {Berger}, \citenamefont {Magrez}, \citenamefont
  {Grioni},\ and\ \citenamefont {Parmigiani}}]{ManzoniPRL15}%
  \BibitemOpen
  \bibfield  {author} {\bibinfo {author} {\bibfnamefont {G.}~\bibnamefont
  {Manzoni}}, \bibinfo {author} {\bibfnamefont {A.}~\bibnamefont {Sterzi}},
  \bibinfo {author} {\bibfnamefont {A.}~\bibnamefont {Crepaldi}}, \bibinfo
  {author} {\bibfnamefont {M.}~\bibnamefont {Diego}}, \bibinfo {author}
  {\bibfnamefont {F.}~\bibnamefont {Cilento}}, \bibinfo {author} {\bibfnamefont
  {M.}~\bibnamefont {Zacchigna}}, \bibinfo {author} {\bibfnamefont
  {P.}~\bibnamefont {Bugnon}}, \bibinfo {author} {\bibfnamefont
  {H.}~\bibnamefont {Berger}}, \bibinfo {author} {\bibfnamefont
  {A.}~\bibnamefont {Magrez}}, \bibinfo {author} {\bibfnamefont
  {M.}~\bibnamefont {Grioni}}, \ and\ \bibinfo {author} {\bibfnamefont
  {F.}~\bibnamefont {Parmigiani}},\ }\href {\doibase
  10.1103/PhysRevLett.115.207402} {\bibfield  {journal} {\bibinfo  {journal}
  {Phys. Rev. Lett.}\ }\textbf {\bibinfo {volume} {115}},\ \bibinfo {pages}
  {207402} (\bibinfo {year} {2015})}\BibitemShut {NoStop}%
\bibitem [{\citenamefont {Moreschini}\ \emph {et~al.}(2016)\citenamefont
  {Moreschini}, \citenamefont {Johannsen}, \citenamefont {Berger},
  \citenamefont {Denlinger}, \citenamefont {Jozwiak}, \citenamefont
  {Rotenberg}, \citenamefont {Kim}, \citenamefont {Bostwick},\ and\
  \citenamefont {Grioni}}]{MoreschiniPRB16}%
  \BibitemOpen
  \bibfield  {author} {\bibinfo {author} {\bibfnamefont {L.}~\bibnamefont
  {Moreschini}}, \bibinfo {author} {\bibfnamefont {J.~C.}\ \bibnamefont
  {Johannsen}}, \bibinfo {author} {\bibfnamefont {H.}~\bibnamefont {Berger}},
  \bibinfo {author} {\bibfnamefont {J.}~\bibnamefont {Denlinger}}, \bibinfo
  {author} {\bibfnamefont {C.}~\bibnamefont {Jozwiak}}, \bibinfo {author}
  {\bibfnamefont {E.}~\bibnamefont {Rotenberg}}, \bibinfo {author}
  {\bibfnamefont {K.~S.}\ \bibnamefont {Kim}}, \bibinfo {author} {\bibfnamefont
  {A.}~\bibnamefont {Bostwick}}, \ and\ \bibinfo {author} {\bibfnamefont
  {M.}~\bibnamefont {Grioni}},\ }\href {\doibase 10.1103/PhysRevB.94.081101}
  {\bibfield  {journal} {\bibinfo  {journal} {Phys. Rev. B}\ }\textbf {\bibinfo
  {volume} {94}},\ \bibinfo {pages} {081101} (\bibinfo {year}
  {2016})}\BibitemShut {NoStop}%
\bibitem [{\citenamefont {Wu}\ \emph {et~al.}(2016)\citenamefont {Wu},
  \citenamefont {Ma}, \citenamefont {Nie}, \citenamefont {Zhao}, \citenamefont
  {Huang}, \citenamefont {Yin}, \citenamefont {Fu}, \citenamefont {Richard},
  \citenamefont {Chen}, \citenamefont {Fang}, \citenamefont {Dai},
  \citenamefont {Weng}, \citenamefont {Qian}, \citenamefont {Ding},\ and\
  \citenamefont {Pan}}]{WuPRX16}%
  \BibitemOpen
  \bibfield  {author} {\bibinfo {author} {\bibfnamefont {R.}~\bibnamefont
  {Wu}}, \bibinfo {author} {\bibfnamefont {J.-Z.}\ \bibnamefont {Ma}}, \bibinfo
  {author} {\bibfnamefont {S.-M.}\ \bibnamefont {Nie}}, \bibinfo {author}
  {\bibfnamefont {L.-X.}\ \bibnamefont {Zhao}}, \bibinfo {author}
  {\bibfnamefont {X.}~\bibnamefont {Huang}}, \bibinfo {author} {\bibfnamefont
  {J.-X.}\ \bibnamefont {Yin}}, \bibinfo {author} {\bibfnamefont {B.-B.}\
  \bibnamefont {Fu}}, \bibinfo {author} {\bibfnamefont {P.}~\bibnamefont
  {Richard}}, \bibinfo {author} {\bibfnamefont {G.-F.}\ \bibnamefont {Chen}},
  \bibinfo {author} {\bibfnamefont {Z.}~\bibnamefont {Fang}}, \bibinfo {author}
  {\bibfnamefont {X.}~\bibnamefont {Dai}}, \bibinfo {author} {\bibfnamefont
  {H.-M.}\ \bibnamefont {Weng}}, \bibinfo {author} {\bibfnamefont
  {T.}~\bibnamefont {Qian}}, \bibinfo {author} {\bibfnamefont {H.}~\bibnamefont
  {Ding}}, \ and\ \bibinfo {author} {\bibfnamefont {S.~H.}\ \bibnamefont
  {Pan}},\ }\href {\doibase 10.1103/PhysRevX.6.021017} {\bibfield  {journal}
  {\bibinfo  {journal} {Phys. Rev. X}\ }\textbf {\bibinfo {volume} {6}},\
  \bibinfo {pages} {021017} (\bibinfo {year} {2016})}\BibitemShut {NoStop}%
\bibitem [{\citenamefont {Manzoni}\ \emph {et~al.}(2016)\citenamefont
  {Manzoni}, \citenamefont {Gragnaniello}, \citenamefont {Aut\`es},
  \citenamefont {Kuhn}, \citenamefont {Sterzi}, \citenamefont {Cilento},
  \citenamefont {Zacchigna}, \citenamefont {Enenkel}, \citenamefont {Vobornik},
  \citenamefont {Barba}, \citenamefont {Bisti}, \citenamefont {Bugnon},
  \citenamefont {Magrez}, \citenamefont {Strocov}, \citenamefont {Berger},
  \citenamefont {Yazyev}, \citenamefont {Fonin}, \citenamefont {Parmigiani},\
  and\ \citenamefont {Crepaldi}}]{ManzoniPRL16}%
  \BibitemOpen
  \bibfield  {author} {\bibinfo {author} {\bibfnamefont {G.}~\bibnamefont
  {Manzoni}}, \bibinfo {author} {\bibfnamefont {L.}~\bibnamefont
  {Gragnaniello}}, \bibinfo {author} {\bibfnamefont {G.}~\bibnamefont
  {Aut\`es}}, \bibinfo {author} {\bibfnamefont {T.}~\bibnamefont {Kuhn}},
  \bibinfo {author} {\bibfnamefont {A.}~\bibnamefont {Sterzi}}, \bibinfo
  {author} {\bibfnamefont {F.}~\bibnamefont {Cilento}}, \bibinfo {author}
  {\bibfnamefont {M.}~\bibnamefont {Zacchigna}}, \bibinfo {author}
  {\bibfnamefont {V.}~\bibnamefont {Enenkel}}, \bibinfo {author} {\bibfnamefont
  {I.}~\bibnamefont {Vobornik}}, \bibinfo {author} {\bibfnamefont
  {L.}~\bibnamefont {Barba}}, \bibinfo {author} {\bibfnamefont
  {F.}~\bibnamefont {Bisti}}, \bibinfo {author} {\bibfnamefont
  {P.}~\bibnamefont {Bugnon}}, \bibinfo {author} {\bibfnamefont
  {A.}~\bibnamefont {Magrez}}, \bibinfo {author} {\bibfnamefont {V.~N.}\
  \bibnamefont {Strocov}}, \bibinfo {author} {\bibfnamefont {H.}~\bibnamefont
  {Berger}}, \bibinfo {author} {\bibfnamefont {O.~V.}\ \bibnamefont {Yazyev}},
  \bibinfo {author} {\bibfnamefont {M.}~\bibnamefont {Fonin}}, \bibinfo
  {author} {\bibfnamefont {F.}~\bibnamefont {Parmigiani}}, \ and\ \bibinfo
  {author} {\bibfnamefont {A.}~\bibnamefont {Crepaldi}},\ }\href {\doibase
  10.1103/PhysRevLett.117.237601} {\bibfield  {journal} {\bibinfo  {journal}
  {Phys. Rev. Lett.}\ }\textbf {\bibinfo {volume} {117}},\ \bibinfo {pages}
  {237601} (\bibinfo {year} {2016})}\BibitemShut {NoStop}%
\bibitem [{\citenamefont {Manzoni}\ \emph {et~al.}(2017)\citenamefont
  {Manzoni}, \citenamefont {Crepaldi}, \citenamefont {Aut{\`e}s}, \citenamefont
  {Sterzi}, \citenamefont {Cilento}, \citenamefont {Akrap}, \citenamefont
  {Vobornik}, \citenamefont {Gragnaniello}, \citenamefont {Bugnon},
  \citenamefont {Fonin} \emph {et~al.}}]{Manzoni17}%
  \BibitemOpen
  \bibfield  {author} {\bibinfo {author} {\bibfnamefont {G.}~\bibnamefont
  {Manzoni}}, \bibinfo {author} {\bibfnamefont {A.}~\bibnamefont {Crepaldi}},
  \bibinfo {author} {\bibfnamefont {G.}~\bibnamefont {Aut{\`e}s}}, \bibinfo
  {author} {\bibfnamefont {A.}~\bibnamefont {Sterzi}}, \bibinfo {author}
  {\bibfnamefont {F.}~\bibnamefont {Cilento}}, \bibinfo {author} {\bibfnamefont
  {A.}~\bibnamefont {Akrap}}, \bibinfo {author} {\bibfnamefont
  {I.}~\bibnamefont {Vobornik}}, \bibinfo {author} {\bibfnamefont
  {L.}~\bibnamefont {Gragnaniello}}, \bibinfo {author} {\bibfnamefont
  {P.}~\bibnamefont {Bugnon}}, \bibinfo {author} {\bibfnamefont
  {M.}~\bibnamefont {Fonin}},  \emph {et~al.},\ }\href@noop {} {\bibfield
  {journal} {\bibinfo  {journal} {Journal of Electron Spectroscopy and Related
  Phenomena}\ }\textbf {\bibinfo {volume} {219}},\ \bibinfo {pages} {9}
  (\bibinfo {year} {2017})}\BibitemShut {NoStop}%
\bibitem [{\citenamefont {Zhang}\ \emph
  {et~al.}(2017{\natexlab{a}})\citenamefont {Zhang}, \citenamefont {Wang},
  \citenamefont {Yu}, \citenamefont {Liu}, \citenamefont {Liang}, \citenamefont
  {Huang}, \citenamefont {Nie}, \citenamefont {Sun}, \citenamefont {Zhang},
  \citenamefont {Shen} \emph {et~al.}}]{ZhangNatComm17}%
  \BibitemOpen
  \bibfield  {author} {\bibinfo {author} {\bibfnamefont {Y.}~\bibnamefont
  {Zhang}}, \bibinfo {author} {\bibfnamefont {C.}~\bibnamefont {Wang}},
  \bibinfo {author} {\bibfnamefont {L.}~\bibnamefont {Yu}}, \bibinfo {author}
  {\bibfnamefont {G.}~\bibnamefont {Liu}}, \bibinfo {author} {\bibfnamefont
  {A.}~\bibnamefont {Liang}}, \bibinfo {author} {\bibfnamefont
  {J.}~\bibnamefont {Huang}}, \bibinfo {author} {\bibfnamefont
  {S.}~\bibnamefont {Nie}}, \bibinfo {author} {\bibfnamefont {X.}~\bibnamefont
  {Sun}}, \bibinfo {author} {\bibfnamefont {Y.}~\bibnamefont {Zhang}}, \bibinfo
  {author} {\bibfnamefont {B.}~\bibnamefont {Shen}},  \emph {et~al.},\
  }\href@noop {} {\bibfield  {journal} {\bibinfo  {journal} {Nature
  communications}\ }\textbf {\bibinfo {volume} {8}},\ \bibinfo {pages} {15512}
  (\bibinfo {year} {2017}{\natexlab{a}})}\BibitemShut {NoStop}%
\bibitem [{\citenamefont {Xiong}\ \emph {et~al.}(2017)\citenamefont {Xiong},
  \citenamefont {Sobota}, \citenamefont {Yang}, \citenamefont {Soifer},
  \citenamefont {Gauthier}, \citenamefont {Lu}, \citenamefont {Lv},
  \citenamefont {Yao}, \citenamefont {Lu}, \citenamefont {Hashimoto},
  \citenamefont {Kirchmann}, \citenamefont {Chen},\ and\ \citenamefont
  {Shen}}]{XiongPRB17}%
  \BibitemOpen
  \bibfield  {author} {\bibinfo {author} {\bibfnamefont {H.}~\bibnamefont
  {Xiong}}, \bibinfo {author} {\bibfnamefont {J.~A.}\ \bibnamefont {Sobota}},
  \bibinfo {author} {\bibfnamefont {S.-L.}\ \bibnamefont {Yang}}, \bibinfo
  {author} {\bibfnamefont {H.}~\bibnamefont {Soifer}}, \bibinfo {author}
  {\bibfnamefont {A.}~\bibnamefont {Gauthier}}, \bibinfo {author}
  {\bibfnamefont {M.-H.}\ \bibnamefont {Lu}}, \bibinfo {author} {\bibfnamefont
  {Y.-Y.}\ \bibnamefont {Lv}}, \bibinfo {author} {\bibfnamefont {S.-H.}\
  \bibnamefont {Yao}}, \bibinfo {author} {\bibfnamefont {D.}~\bibnamefont
  {Lu}}, \bibinfo {author} {\bibfnamefont {M.}~\bibnamefont {Hashimoto}},
  \bibinfo {author} {\bibfnamefont {P.~S.}\ \bibnamefont {Kirchmann}}, \bibinfo
  {author} {\bibfnamefont {Y.-F.}\ \bibnamefont {Chen}}, \ and\ \bibinfo
  {author} {\bibfnamefont {Z.-X.}\ \bibnamefont {Shen}},\ }\href {\doibase
  10.1103/PhysRevB.95.195119} {\bibfield  {journal} {\bibinfo  {journal} {Phys.
  Rev. B}\ }\textbf {\bibinfo {volume} {95}},\ \bibinfo {pages} {195119}
  (\bibinfo {year} {2017})}\BibitemShut {NoStop}%
\bibitem [{\citenamefont {Li}\ \emph {et~al.}(2016{\natexlab{a}})\citenamefont
  {Li}, \citenamefont {Huang}, \citenamefont {Lv}, \citenamefont {Zhang},
  \citenamefont {Yang}, \citenamefont {Zhang}, \citenamefont {Chen},
  \citenamefont {Yao}, \citenamefont {Zhou}, \citenamefont {Lu}, \citenamefont
  {Sheng}, \citenamefont {Li}, \citenamefont {Jia}, \citenamefont {Xue},
  \citenamefont {Chen},\ and\ \citenamefont {Xing}}]{LiPRL16}%
  \BibitemOpen
  \bibfield  {author} {\bibinfo {author} {\bibfnamefont {X.-B.}\ \bibnamefont
  {Li}}, \bibinfo {author} {\bibfnamefont {W.-K.}\ \bibnamefont {Huang}},
  \bibinfo {author} {\bibfnamefont {Y.-Y.}\ \bibnamefont {Lv}}, \bibinfo
  {author} {\bibfnamefont {K.-W.}\ \bibnamefont {Zhang}}, \bibinfo {author}
  {\bibfnamefont {C.-L.}\ \bibnamefont {Yang}}, \bibinfo {author}
  {\bibfnamefont {B.-B.}\ \bibnamefont {Zhang}}, \bibinfo {author}
  {\bibfnamefont {Y.~B.}\ \bibnamefont {Chen}}, \bibinfo {author}
  {\bibfnamefont {S.-H.}\ \bibnamefont {Yao}}, \bibinfo {author} {\bibfnamefont
  {J.}~\bibnamefont {Zhou}}, \bibinfo {author} {\bibfnamefont {M.-H.}\
  \bibnamefont {Lu}}, \bibinfo {author} {\bibfnamefont {L.}~\bibnamefont
  {Sheng}}, \bibinfo {author} {\bibfnamefont {S.-C.}\ \bibnamefont {Li}},
  \bibinfo {author} {\bibfnamefont {J.-F.}\ \bibnamefont {Jia}}, \bibinfo
  {author} {\bibfnamefont {Q.-K.}\ \bibnamefont {Xue}}, \bibinfo {author}
  {\bibfnamefont {Y.-F.}\ \bibnamefont {Chen}}, \ and\ \bibinfo {author}
  {\bibfnamefont {D.-Y.}\ \bibnamefont {Xing}},\ }\href {\doibase
  10.1103/PhysRevLett.116.176803} {\bibfield  {journal} {\bibinfo  {journal}
  {Phys. Rev. Lett.}\ }\textbf {\bibinfo {volume} {116}},\ \bibinfo {pages}
  {176803} (\bibinfo {year} {2016}{\natexlab{a}})}\BibitemShut {NoStop}%
\bibitem [{\citenamefont {Chen}\ \emph
  {et~al.}(2015{\natexlab{a}})\citenamefont {Chen}, \citenamefont {Zhang},
  \citenamefont {Schneeloch}, \citenamefont {Zhang}, \citenamefont {Li},
  \citenamefont {Gu},\ and\ \citenamefont {Wang}}]{WangPRB15}%
  \BibitemOpen
  \bibfield  {author} {\bibinfo {author} {\bibfnamefont {R.~Y.}\ \bibnamefont
  {Chen}}, \bibinfo {author} {\bibfnamefont {S.~J.}\ \bibnamefont {Zhang}},
  \bibinfo {author} {\bibfnamefont {J.~A.}\ \bibnamefont {Schneeloch}},
  \bibinfo {author} {\bibfnamefont {C.}~\bibnamefont {Zhang}}, \bibinfo
  {author} {\bibfnamefont {Q.}~\bibnamefont {Li}}, \bibinfo {author}
  {\bibfnamefont {G.~D.}\ \bibnamefont {Gu}}, \ and\ \bibinfo {author}
  {\bibfnamefont {N.~L.}\ \bibnamefont {Wang}},\ }\href {\doibase
  10.1103/PhysRevB.92.075107} {\bibfield  {journal} {\bibinfo  {journal} {Phys.
  Rev. B}\ }\textbf {\bibinfo {volume} {92}},\ \bibinfo {pages} {075107}
  (\bibinfo {year} {2015}{\natexlab{a}})}\BibitemShut {NoStop}%
\bibitem [{\citenamefont {Chen}\ \emph
  {et~al.}(2015{\natexlab{b}})\citenamefont {Chen}, \citenamefont {Chen},
  \citenamefont {Song}, \citenamefont {Schneeloch}, \citenamefont {Gu},
  \citenamefont {Wang},\ and\ \citenamefont {Wang}}]{WangPRL15}%
  \BibitemOpen
  \bibfield  {author} {\bibinfo {author} {\bibfnamefont {R.~Y.}\ \bibnamefont
  {Chen}}, \bibinfo {author} {\bibfnamefont {Z.~G.}\ \bibnamefont {Chen}},
  \bibinfo {author} {\bibfnamefont {X.-Y.}\ \bibnamefont {Song}}, \bibinfo
  {author} {\bibfnamefont {J.~A.}\ \bibnamefont {Schneeloch}}, \bibinfo
  {author} {\bibfnamefont {G.~D.}\ \bibnamefont {Gu}}, \bibinfo {author}
  {\bibfnamefont {F.}~\bibnamefont {Wang}}, \ and\ \bibinfo {author}
  {\bibfnamefont {N.~L.}\ \bibnamefont {Wang}},\ }\href {\doibase
  10.1103/PhysRevLett.115.176404} {\bibfield  {journal} {\bibinfo  {journal}
  {Phys. Rev. Lett.}\ }\textbf {\bibinfo {volume} {115}},\ \bibinfo {pages}
  {176404} (\bibinfo {year} {2015}{\natexlab{b}})}\BibitemShut {NoStop}%
\bibitem [{\citenamefont {Jiang}\ \emph {et~al.}(2017)\citenamefont {Jiang},
  \citenamefont {Dun}, \citenamefont {Zhou}, \citenamefont {Lu}, \citenamefont
  {Chen}, \citenamefont {Moon}, \citenamefont {Besara}, \citenamefont
  {Siegrist}, \citenamefont {Baumbach}, \citenamefont {Smirnov},\ and\
  \citenamefont {Jiang}}]{JiangPRB17}%
  \BibitemOpen
  \bibfield  {author} {\bibinfo {author} {\bibfnamefont {Y.}~\bibnamefont
  {Jiang}}, \bibinfo {author} {\bibfnamefont {Z.~L.}\ \bibnamefont {Dun}},
  \bibinfo {author} {\bibfnamefont {H.~D.}\ \bibnamefont {Zhou}}, \bibinfo
  {author} {\bibfnamefont {Z.}~\bibnamefont {Lu}}, \bibinfo {author}
  {\bibfnamefont {K.-W.}\ \bibnamefont {Chen}}, \bibinfo {author}
  {\bibfnamefont {S.}~\bibnamefont {Moon}}, \bibinfo {author} {\bibfnamefont
  {T.}~\bibnamefont {Besara}}, \bibinfo {author} {\bibfnamefont {T.~M.}\
  \bibnamefont {Siegrist}}, \bibinfo {author} {\bibfnamefont {R.~E.}\
  \bibnamefont {Baumbach}}, \bibinfo {author} {\bibfnamefont {D.}~\bibnamefont
  {Smirnov}}, \ and\ \bibinfo {author} {\bibfnamefont {Z.}~\bibnamefont
  {Jiang}},\ }\href {\doibase 10.1103/PhysRevB.96.041101} {\bibfield  {journal}
  {\bibinfo  {journal} {Phys. Rev. B}\ }\textbf {\bibinfo {volume} {96}},\
  \bibinfo {pages} {041101} (\bibinfo {year} {2017})}\BibitemShut {NoStop}%
\bibitem [{\citenamefont {Chen}\ \emph {et~al.}(2017)\citenamefont {Chen},
  \citenamefont {Chen}, \citenamefont {Zhong}, \citenamefont {Schneeloch},
  \citenamefont {Zhang}, \citenamefont {Huang}, \citenamefont {Qu},
  \citenamefont {Yu}, \citenamefont {Li}, \citenamefont {Gu},\ and\
  \citenamefont {Wang}}]{WangPNAS17}%
  \BibitemOpen
  \bibfield  {author} {\bibinfo {author} {\bibfnamefont {Z.-G.}\ \bibnamefont
  {Chen}}, \bibinfo {author} {\bibfnamefont {R.~Y.}\ \bibnamefont {Chen}},
  \bibinfo {author} {\bibfnamefont {R.~D.}\ \bibnamefont {Zhong}}, \bibinfo
  {author} {\bibfnamefont {J.}~\bibnamefont {Schneeloch}}, \bibinfo {author}
  {\bibfnamefont {C.}~\bibnamefont {Zhang}}, \bibinfo {author} {\bibfnamefont
  {Y.}~\bibnamefont {Huang}}, \bibinfo {author} {\bibfnamefont
  {F.}~\bibnamefont {Qu}}, \bibinfo {author} {\bibfnamefont {R.}~\bibnamefont
  {Yu}}, \bibinfo {author} {\bibfnamefont {Q.}~\bibnamefont {Li}}, \bibinfo
  {author} {\bibfnamefont {G.~D.}\ \bibnamefont {Gu}}, \ and\ \bibinfo {author}
  {\bibfnamefont {N.~L.}\ \bibnamefont {Wang}},\ }\href {\doibase
  10.1073/pnas.1613110114} {\bibfield  {journal} {\bibinfo  {journal}
  {Proceedings of the National Academy of Sciences}\ }\textbf {\bibinfo
  {volume} {114}},\ \bibinfo {pages} {816} (\bibinfo {year} {2017})},\ \Eprint
  {http://arxiv.org/abs/https://www.pnas.org/content/114/5/816.full.pdf}
  {https://www.pnas.org/content/114/5/816.full.pdf} \BibitemShut {NoStop}%
\bibitem [{\citenamefont {Zheng}\ \emph {et~al.}(2016)\citenamefont {Zheng},
  \citenamefont {Lu}, \citenamefont {Zhu}, \citenamefont {Ning}, \citenamefont
  {Han}, \citenamefont {Zhang}, \citenamefont {Zhang}, \citenamefont {Xi},
  \citenamefont {Yang}, \citenamefont {Du}, \citenamefont {Yang}, \citenamefont
  {Zhang},\ and\ \citenamefont {Tian}}]{TianPRB16}%
  \BibitemOpen
  \bibfield  {author} {\bibinfo {author} {\bibfnamefont {G.}~\bibnamefont
  {Zheng}}, \bibinfo {author} {\bibfnamefont {J.}~\bibnamefont {Lu}}, \bibinfo
  {author} {\bibfnamefont {X.}~\bibnamefont {Zhu}}, \bibinfo {author}
  {\bibfnamefont {W.}~\bibnamefont {Ning}}, \bibinfo {author} {\bibfnamefont
  {Y.}~\bibnamefont {Han}}, \bibinfo {author} {\bibfnamefont {H.}~\bibnamefont
  {Zhang}}, \bibinfo {author} {\bibfnamefont {J.}~\bibnamefont {Zhang}},
  \bibinfo {author} {\bibfnamefont {C.}~\bibnamefont {Xi}}, \bibinfo {author}
  {\bibfnamefont {J.}~\bibnamefont {Yang}}, \bibinfo {author} {\bibfnamefont
  {H.}~\bibnamefont {Du}}, \bibinfo {author} {\bibfnamefont {K.}~\bibnamefont
  {Yang}}, \bibinfo {author} {\bibfnamefont {Y.}~\bibnamefont {Zhang}}, \ and\
  \bibinfo {author} {\bibfnamefont {M.}~\bibnamefont {Tian}},\ }\href {\doibase
  10.1103/PhysRevB.93.115414} {\bibfield  {journal} {\bibinfo  {journal} {Phys.
  Rev. B}\ }\textbf {\bibinfo {volume} {93}},\ \bibinfo {pages} {115414}
  (\bibinfo {year} {2016})}\BibitemShut {NoStop}%
\bibitem [{\citenamefont {Liu}\ \emph {et~al.}(2016)\citenamefont {Liu},
  \citenamefont {Yuan}, \citenamefont {Zhang}, \citenamefont {Jin},
  \citenamefont {Narayan}, \citenamefont {Luo}, \citenamefont {Chen},
  \citenamefont {Yang}, \citenamefont {Zou}, \citenamefont {Wu} \emph
  {et~al.}}]{XiuNatComm16}%
  \BibitemOpen
  \bibfield  {author} {\bibinfo {author} {\bibfnamefont {Y.}~\bibnamefont
  {Liu}}, \bibinfo {author} {\bibfnamefont {X.}~\bibnamefont {Yuan}}, \bibinfo
  {author} {\bibfnamefont {C.}~\bibnamefont {Zhang}}, \bibinfo {author}
  {\bibfnamefont {Z.}~\bibnamefont {Jin}}, \bibinfo {author} {\bibfnamefont
  {A.}~\bibnamefont {Narayan}}, \bibinfo {author} {\bibfnamefont
  {C.}~\bibnamefont {Luo}}, \bibinfo {author} {\bibfnamefont {Z.}~\bibnamefont
  {Chen}}, \bibinfo {author} {\bibfnamefont {L.}~\bibnamefont {Yang}}, \bibinfo
  {author} {\bibfnamefont {J.}~\bibnamefont {Zou}}, \bibinfo {author}
  {\bibfnamefont {X.}~\bibnamefont {Wu}},  \emph {et~al.},\ }\href@noop {}
  {\bibfield  {journal} {\bibinfo  {journal} {Nature communications}\ }\textbf
  {\bibinfo {volume} {7}},\ \bibinfo {pages} {12516} (\bibinfo {year}
  {2016})}\BibitemShut {NoStop}%
\bibitem [{\citenamefont {Zheng}\ \emph {et~al.}(2017)\citenamefont {Zheng},
  \citenamefont {Zhu}, \citenamefont {Liu}, \citenamefont {Lu}, \citenamefont
  {Ning}, \citenamefont {Zhang}, \citenamefont {Gao}, \citenamefont {Han},
  \citenamefont {Yang}, \citenamefont {Du}, \citenamefont {Yang}, \citenamefont
  {Zhang},\ and\ \citenamefont {Tian}}]{TianPRB17}%
  \BibitemOpen
  \bibfield  {author} {\bibinfo {author} {\bibfnamefont {G.}~\bibnamefont
  {Zheng}}, \bibinfo {author} {\bibfnamefont {X.}~\bibnamefont {Zhu}}, \bibinfo
  {author} {\bibfnamefont {Y.}~\bibnamefont {Liu}}, \bibinfo {author}
  {\bibfnamefont {J.}~\bibnamefont {Lu}}, \bibinfo {author} {\bibfnamefont
  {W.}~\bibnamefont {Ning}}, \bibinfo {author} {\bibfnamefont {H.}~\bibnamefont
  {Zhang}}, \bibinfo {author} {\bibfnamefont {W.}~\bibnamefont {Gao}}, \bibinfo
  {author} {\bibfnamefont {Y.}~\bibnamefont {Han}}, \bibinfo {author}
  {\bibfnamefont {J.}~\bibnamefont {Yang}}, \bibinfo {author} {\bibfnamefont
  {H.}~\bibnamefont {Du}}, \bibinfo {author} {\bibfnamefont {K.}~\bibnamefont
  {Yang}}, \bibinfo {author} {\bibfnamefont {Y.}~\bibnamefont {Zhang}}, \ and\
  \bibinfo {author} {\bibfnamefont {M.}~\bibnamefont {Tian}},\ }\href {\doibase
  10.1103/PhysRevB.96.121401} {\bibfield  {journal} {\bibinfo  {journal} {Phys.
  Rev. B}\ }\textbf {\bibinfo {volume} {96}},\ \bibinfo {pages} {121401}
  (\bibinfo {year} {2017})}\BibitemShut {NoStop}%
\bibitem [{\citenamefont {Zhang}\ \emph
  {et~al.}(2017{\natexlab{b}})\citenamefont {Zhang}, \citenamefont {Guo},
  \citenamefont {Zhu}, \citenamefont {Ma}, \citenamefont {Zheng}, \citenamefont
  {Wang}, \citenamefont {Pi}, \citenamefont {Chen}, \citenamefont {Yuan},\ and\
  \citenamefont {Tian}}]{TianPRL17}%
  \BibitemOpen
  \bibfield  {author} {\bibinfo {author} {\bibfnamefont {J.~L.}\ \bibnamefont
  {Zhang}}, \bibinfo {author} {\bibfnamefont {C.~Y.}\ \bibnamefont {Guo}},
  \bibinfo {author} {\bibfnamefont {X.~D.}\ \bibnamefont {Zhu}}, \bibinfo
  {author} {\bibfnamefont {L.}~\bibnamefont {Ma}}, \bibinfo {author}
  {\bibfnamefont {G.~L.}\ \bibnamefont {Zheng}}, \bibinfo {author}
  {\bibfnamefont {Y.~Q.}\ \bibnamefont {Wang}}, \bibinfo {author}
  {\bibfnamefont {L.}~\bibnamefont {Pi}}, \bibinfo {author} {\bibfnamefont
  {Y.}~\bibnamefont {Chen}}, \bibinfo {author} {\bibfnamefont {H.~Q.}\
  \bibnamefont {Yuan}}, \ and\ \bibinfo {author} {\bibfnamefont {M.~L.}\
  \bibnamefont {Tian}},\ }\href {\doibase 10.1103/PhysRevLett.118.206601}
  {\bibfield  {journal} {\bibinfo  {journal} {Phys. Rev. Lett.}\ }\textbf
  {\bibinfo {volume} {118}},\ \bibinfo {pages} {206601} (\bibinfo {year}
  {2017}{\natexlab{b}})}\BibitemShut {NoStop}%
\bibitem [{\citenamefont {Li}\ \emph {et~al.}(2016{\natexlab{b}})\citenamefont
  {Li}, \citenamefont {Kharzeev}, \citenamefont {Zhang}, \citenamefont {Huang},
  \citenamefont {Pletikosi{\'c}}, \citenamefont {Fedorov}, \citenamefont
  {Zhong}, \citenamefont {Schneeloch}, \citenamefont {Gu},\ and\ \citenamefont
  {Valla}}]{LiNatPhy16}%
  \BibitemOpen
  \bibfield  {author} {\bibinfo {author} {\bibfnamefont {Q.}~\bibnamefont
  {Li}}, \bibinfo {author} {\bibfnamefont {D.~E.}\ \bibnamefont {Kharzeev}},
  \bibinfo {author} {\bibfnamefont {C.}~\bibnamefont {Zhang}}, \bibinfo
  {author} {\bibfnamefont {Y.}~\bibnamefont {Huang}}, \bibinfo {author}
  {\bibfnamefont {I.}~\bibnamefont {Pletikosi{\'c}}}, \bibinfo {author}
  {\bibfnamefont {A.}~\bibnamefont {Fedorov}}, \bibinfo {author} {\bibfnamefont
  {R.}~\bibnamefont {Zhong}}, \bibinfo {author} {\bibfnamefont
  {J.}~\bibnamefont {Schneeloch}}, \bibinfo {author} {\bibfnamefont
  {G.}~\bibnamefont {Gu}}, \ and\ \bibinfo {author} {\bibfnamefont
  {T.}~\bibnamefont {Valla}},\ }\href@noop {} {\bibfield  {journal} {\bibinfo
  {journal} {Nature Physics}\ }\textbf {\bibinfo {volume} {12}},\ \bibinfo
  {pages} {550} (\bibinfo {year} {2016}{\natexlab{b}})}\BibitemShut {NoStop}%
\bibitem [{\citenamefont {Liang}\ \emph {et~al.}(2018)\citenamefont {Liang},
  \citenamefont {Lin}, \citenamefont {Gibson}, \citenamefont {Kushwaha},
  \citenamefont {Liu}, \citenamefont {Wang}, \citenamefont {Xiong},
  \citenamefont {Sobota}, \citenamefont {Hashimoto}, \citenamefont {Kirchmann}
  \emph {et~al.}}]{LiangNatPhy18}%
  \BibitemOpen
  \bibfield  {author} {\bibinfo {author} {\bibfnamefont {T.}~\bibnamefont
  {Liang}}, \bibinfo {author} {\bibfnamefont {J.}~\bibnamefont {Lin}}, \bibinfo
  {author} {\bibfnamefont {Q.}~\bibnamefont {Gibson}}, \bibinfo {author}
  {\bibfnamefont {S.}~\bibnamefont {Kushwaha}}, \bibinfo {author}
  {\bibfnamefont {M.}~\bibnamefont {Liu}}, \bibinfo {author} {\bibfnamefont
  {W.}~\bibnamefont {Wang}}, \bibinfo {author} {\bibfnamefont {H.}~\bibnamefont
  {Xiong}}, \bibinfo {author} {\bibfnamefont {J.~A.}\ \bibnamefont {Sobota}},
  \bibinfo {author} {\bibfnamefont {M.}~\bibnamefont {Hashimoto}}, \bibinfo
  {author} {\bibfnamefont {P.~S.}\ \bibnamefont {Kirchmann}},  \emph {et~al.},\
  }\href@noop {} {\bibfield  {journal} {\bibinfo  {journal} {Nature Physics}\
  }\textbf {\bibinfo {volume} {14}},\ \bibinfo {pages} {451} (\bibinfo {year}
  {2018})}\BibitemShut {NoStop}%
\bibitem [{\citenamefont {Wang}\ \emph {et~al.}(2018)\citenamefont {Wang},
  \citenamefont {Liu}, \citenamefont {Li}, \citenamefont {Liu}, \citenamefont
  {Wang}, \citenamefont {Liu}, \citenamefont {Dai}, \citenamefont {Wang},
  \citenamefont {Li}, \citenamefont {Yan} \emph {et~al.}}]{WangSciAdv18}%
  \BibitemOpen
  \bibfield  {author} {\bibinfo {author} {\bibfnamefont {H.}~\bibnamefont
  {Wang}}, \bibinfo {author} {\bibfnamefont {H.}~\bibnamefont {Liu}}, \bibinfo
  {author} {\bibfnamefont {Y.}~\bibnamefont {Li}}, \bibinfo {author}
  {\bibfnamefont {Y.}~\bibnamefont {Liu}}, \bibinfo {author} {\bibfnamefont
  {J.}~\bibnamefont {Wang}}, \bibinfo {author} {\bibfnamefont {J.}~\bibnamefont
  {Liu}}, \bibinfo {author} {\bibfnamefont {J.-Y.}\ \bibnamefont {Dai}},
  \bibinfo {author} {\bibfnamefont {Y.}~\bibnamefont {Wang}}, \bibinfo {author}
  {\bibfnamefont {L.}~\bibnamefont {Li}}, \bibinfo {author} {\bibfnamefont
  {J.}~\bibnamefont {Yan}},  \emph {et~al.},\ }\href@noop {} {\bibfield
  {journal} {\bibinfo  {journal} {Science advances}\ }\textbf {\bibinfo
  {volume} {4}},\ \bibinfo {pages} {eaau5096} (\bibinfo {year}
  {2018})}\BibitemShut {NoStop}%
\bibitem [{\citenamefont {Tang}\ \emph {et~al.}(2019)\citenamefont {Tang},
  \citenamefont {Ren}, \citenamefont {Wang}, \citenamefont {Zhong},
  \citenamefont {Schneeloch}, \citenamefont {Yang}, \citenamefont {Yang},
  \citenamefont {Lee}, \citenamefont {Gu}, \citenamefont {Qiao} \emph
  {et~al.}}]{TangNat19}%
  \BibitemOpen
  \bibfield  {author} {\bibinfo {author} {\bibfnamefont {F.}~\bibnamefont
  {Tang}}, \bibinfo {author} {\bibfnamefont {Y.}~\bibnamefont {Ren}}, \bibinfo
  {author} {\bibfnamefont {P.}~\bibnamefont {Wang}}, \bibinfo {author}
  {\bibfnamefont {R.}~\bibnamefont {Zhong}}, \bibinfo {author} {\bibfnamefont
  {J.}~\bibnamefont {Schneeloch}}, \bibinfo {author} {\bibfnamefont {S.~A.}\
  \bibnamefont {Yang}}, \bibinfo {author} {\bibfnamefont {K.}~\bibnamefont
  {Yang}}, \bibinfo {author} {\bibfnamefont {P.~A.}\ \bibnamefont {Lee}},
  \bibinfo {author} {\bibfnamefont {G.}~\bibnamefont {Gu}}, \bibinfo {author}
  {\bibfnamefont {Z.}~\bibnamefont {Qiao}},  \emph {et~al.},\ }\href@noop {}
  {\bibfield  {journal} {\bibinfo  {journal} {Nature}\ }\textbf {\bibinfo
  {volume} {569}},\ \bibinfo {pages} {537} (\bibinfo {year}
  {2019})}\BibitemShut {NoStop}%
\bibitem [{\citenamefont {Zhang}\ \emph {et~al.}(2020)\citenamefont {Zhang},
  \citenamefont {Wang}, \citenamefont {Skinner}, \citenamefont {Bi},
  \citenamefont {Kozii}, \citenamefont {Cho}, \citenamefont {Zhong},
  \citenamefont {Schneeloch}, \citenamefont {Yu}, \citenamefont {Gu} \emph
  {et~al.}}]{Zhang2020}%
  \BibitemOpen
  \bibfield  {author} {\bibinfo {author} {\bibfnamefont {W.}~\bibnamefont
  {Zhang}}, \bibinfo {author} {\bibfnamefont {P.}~\bibnamefont {Wang}},
  \bibinfo {author} {\bibfnamefont {B.}~\bibnamefont {Skinner}}, \bibinfo
  {author} {\bibfnamefont {R.}~\bibnamefont {Bi}}, \bibinfo {author}
  {\bibfnamefont {V.}~\bibnamefont {Kozii}}, \bibinfo {author} {\bibfnamefont
  {C.-W.}\ \bibnamefont {Cho}}, \bibinfo {author} {\bibfnamefont
  {R.}~\bibnamefont {Zhong}}, \bibinfo {author} {\bibfnamefont
  {J.}~\bibnamefont {Schneeloch}}, \bibinfo {author} {\bibfnamefont
  {D.}~\bibnamefont {Yu}}, \bibinfo {author} {\bibfnamefont {G.}~\bibnamefont
  {Gu}},  \emph {et~al.},\ }\href@noop {} {\bibfield  {journal} {\bibinfo
  {journal} {Nature communications}\ }\textbf {\bibinfo {volume} {11}},\
  \bibinfo {pages} {1} (\bibinfo {year} {2020})}\BibitemShut {NoStop}%
\bibitem [{\citenamefont {Lv}\ \emph {et~al.}(2017)\citenamefont {Lv},
  \citenamefont {Zhang}, \citenamefont {Zhang}, \citenamefont {Pang},
  \citenamefont {Yao}, \citenamefont {Chen}, \citenamefont {Ye}, \citenamefont
  {Zhou}, \citenamefont {Zhang},\ and\ \citenamefont {Chen}}]{Chen17}%
  \BibitemOpen
  \bibfield  {author} {\bibinfo {author} {\bibfnamefont {Y.-Y.}\ \bibnamefont
  {Lv}}, \bibinfo {author} {\bibfnamefont {F.}~\bibnamefont {Zhang}}, \bibinfo
  {author} {\bibfnamefont {B.-B.}\ \bibnamefont {Zhang}}, \bibinfo {author}
  {\bibfnamefont {B.}~\bibnamefont {Pang}}, \bibinfo {author} {\bibfnamefont
  {S.-H.}\ \bibnamefont {Yao}}, \bibinfo {author} {\bibfnamefont
  {Y.}~\bibnamefont {Chen}}, \bibinfo {author} {\bibfnamefont {L.}~\bibnamefont
  {Ye}}, \bibinfo {author} {\bibfnamefont {J.}~\bibnamefont {Zhou}}, \bibinfo
  {author} {\bibfnamefont {S.-T.}\ \bibnamefont {Zhang}}, \ and\ \bibinfo
  {author} {\bibfnamefont {Y.-F.}\ \bibnamefont {Chen}},\ }\href {\doibase
  https://doi.org/10.1016/j.jcrysgro.2016.04.042} {\bibfield  {journal}
  {\bibinfo  {journal} {Journal of Crystal Growth}\ }\textbf {\bibinfo {volume}
  {457}},\ \bibinfo {pages} {250 } (\bibinfo {year} {2017})}\BibitemShut
  {NoStop}%
\bibitem [{\citenamefont {Mutch}\ \emph {et~al.}(2019)\citenamefont {Mutch},
  \citenamefont {Chen}, \citenamefont {Went}, \citenamefont {Qian},
  \citenamefont {Wilson}, \citenamefont {Andreev}, \citenamefont {Chen},\ and\
  \citenamefont {Chu}}]{Chu19}%
  \BibitemOpen
  \bibfield  {author} {\bibinfo {author} {\bibfnamefont {J.}~\bibnamefont
  {Mutch}}, \bibinfo {author} {\bibfnamefont {W.-C.}\ \bibnamefont {Chen}},
  \bibinfo {author} {\bibfnamefont {P.}~\bibnamefont {Went}}, \bibinfo {author}
  {\bibfnamefont {T.}~\bibnamefont {Qian}}, \bibinfo {author} {\bibfnamefont
  {I.~Z.}\ \bibnamefont {Wilson}}, \bibinfo {author} {\bibfnamefont
  {A.}~\bibnamefont {Andreev}}, \bibinfo {author} {\bibfnamefont {C.-C.}\
  \bibnamefont {Chen}}, \ and\ \bibinfo {author} {\bibfnamefont {J.-H.}\
  \bibnamefont {Chu}},\ }\href@noop {} {\bibfield  {journal} {\bibinfo
  {journal} {Science advances}\ }\textbf {\bibinfo {volume} {5}},\ \bibinfo
  {pages} {eaav9771} (\bibinfo {year} {2019})}\BibitemShut {NoStop}%
\bibitem [{\citenamefont {McIlroy}\ \emph {et~al.}(2004)\citenamefont
  {McIlroy}, \citenamefont {Moore}, \citenamefont {Zhang}, \citenamefont
  {Wharton}, \citenamefont {Kempton}, \citenamefont {Littleton}, \citenamefont
  {Wilson}, \citenamefont {Tritt},\ and\ \citenamefont {Olson}}]{Mcilroy04}%
  \BibitemOpen
  \bibfield  {author} {\bibinfo {author} {\bibfnamefont {D.}~\bibnamefont
  {McIlroy}}, \bibinfo {author} {\bibfnamefont {S.}~\bibnamefont {Moore}},
  \bibinfo {author} {\bibfnamefont {D.}~\bibnamefont {Zhang}}, \bibinfo
  {author} {\bibfnamefont {J.}~\bibnamefont {Wharton}}, \bibinfo {author}
  {\bibfnamefont {B.}~\bibnamefont {Kempton}}, \bibinfo {author} {\bibfnamefont
  {R.}~\bibnamefont {Littleton}}, \bibinfo {author} {\bibfnamefont
  {M.}~\bibnamefont {Wilson}}, \bibinfo {author} {\bibfnamefont
  {T.}~\bibnamefont {Tritt}}, \ and\ \bibinfo {author} {\bibfnamefont
  {C.}~\bibnamefont {Olson}},\ }\href@noop {} {\bibfield  {journal} {\bibinfo
  {journal} {Journal of Physics: Condensed Matter}\ }\textbf {\bibinfo {volume}
  {16}},\ \bibinfo {pages} {L359} (\bibinfo {year} {2004})}\BibitemShut
  {NoStop}%
\bibitem [{\citenamefont {Zhou}\ \emph {et~al.}(2016)\citenamefont {Zhou},
  \citenamefont {Wu}, \citenamefont {Ning}, \citenamefont {Li}, \citenamefont
  {Du}, \citenamefont {Chen}, \citenamefont {Zhang}, \citenamefont {Chi},
  \citenamefont {Wang}, \citenamefont {Zhu}, \citenamefont {Lu}, \citenamefont
  {Ji}, \citenamefont {Wan}, \citenamefont {Yang}, \citenamefont {Sun},
  \citenamefont {Yang}, \citenamefont {Tian}, \citenamefont {Zhang},\ and\
  \citenamefont {Mao}}]{ZhouPNAS16}%
  \BibitemOpen
  \bibfield  {author} {\bibinfo {author} {\bibfnamefont {Y.}~\bibnamefont
  {Zhou}}, \bibinfo {author} {\bibfnamefont {J.}~\bibnamefont {Wu}}, \bibinfo
  {author} {\bibfnamefont {W.}~\bibnamefont {Ning}}, \bibinfo {author}
  {\bibfnamefont {N.}~\bibnamefont {Li}}, \bibinfo {author} {\bibfnamefont
  {Y.}~\bibnamefont {Du}}, \bibinfo {author} {\bibfnamefont {X.}~\bibnamefont
  {Chen}}, \bibinfo {author} {\bibfnamefont {R.}~\bibnamefont {Zhang}},
  \bibinfo {author} {\bibfnamefont {Z.}~\bibnamefont {Chi}}, \bibinfo {author}
  {\bibfnamefont {X.}~\bibnamefont {Wang}}, \bibinfo {author} {\bibfnamefont
  {X.}~\bibnamefont {Zhu}}, \bibinfo {author} {\bibfnamefont {P.}~\bibnamefont
  {Lu}}, \bibinfo {author} {\bibfnamefont {C.}~\bibnamefont {Ji}}, \bibinfo
  {author} {\bibfnamefont {X.}~\bibnamefont {Wan}}, \bibinfo {author}
  {\bibfnamefont {Z.}~\bibnamefont {Yang}}, \bibinfo {author} {\bibfnamefont
  {J.}~\bibnamefont {Sun}}, \bibinfo {author} {\bibfnamefont {W.}~\bibnamefont
  {Yang}}, \bibinfo {author} {\bibfnamefont {M.}~\bibnamefont {Tian}}, \bibinfo
  {author} {\bibfnamefont {Y.}~\bibnamefont {Zhang}}, \ and\ \bibinfo {author}
  {\bibfnamefont {H.-k.}\ \bibnamefont {Mao}},\ }\href {\doibase
  10.1073/pnas.1601262113} {\bibfield  {journal} {\bibinfo  {journal}
  {Proceedings of the National Academy of Sciences}\ }\textbf {\bibinfo
  {volume} {113}},\ \bibinfo {pages} {2904} (\bibinfo {year} {2016})},\ \Eprint
  {http://arxiv.org/abs/https://www.pnas.org/content/113/11/2904.full.pdf}
  {https://www.pnas.org/content/113/11/2904.full.pdf} \BibitemShut {NoStop}%
\bibitem [{SI()}]{SI}%
  \BibitemOpen
  \href@noop {} {\bibinfo  {journal} {See Supplementary Materials}\
  }\BibitemShut {NoStop}%
\bibitem [{\citenamefont {Kittel}\ \emph {et~al.}(1996)\citenamefont {Kittel},
  \citenamefont {McEuen},\ and\ \citenamefont {McEuen}}]{Kittel}%
  \BibitemOpen
\bibfield  {journal} {  }\bibfield  {author} {\bibinfo {author} {\bibfnamefont
  {C.}~\bibnamefont {Kittel}}, \bibinfo {author} {\bibfnamefont
  {P.}~\bibnamefont {McEuen}}, \ and\ \bibinfo {author} {\bibfnamefont
  {P.}~\bibnamefont {McEuen}},\ }\href@noop {} {\emph {\bibinfo {title}
  {Introduction to solid state physics}}},\ Vol.~\bibinfo {volume} {8}\
  (\bibinfo  {publisher} {Wiley New York},\ \bibinfo {year} {1996})\BibitemShut
  {NoStop}%
\bibitem [{\citenamefont {Shoenberg}(1984)}]{Shoenberg}%
  \BibitemOpen
  \bibfield  {author} {\bibinfo {author} {\bibfnamefont {D.}~\bibnamefont
  {Shoenberg}},\ }\href {https://books.google.com/books?id=GO7cNJl2juAC} {\emph
  {\bibinfo {title} {Magnetic Oscillations in Metals}}},\ Arnold and Caroline
  Rose Monograph Series of the American So\ (\bibinfo  {publisher} {Cambridge
  University Press},\ \bibinfo {year} {1984})\BibitemShut {NoStop}%
\bibitem [{\citenamefont {Skinner}\ and\ \citenamefont {Fu}(2018)}]{Liang18}%
  \BibitemOpen
  \bibfield  {author} {\bibinfo {author} {\bibfnamefont {B.}~\bibnamefont
  {Skinner}}\ and\ \bibinfo {author} {\bibfnamefont {L.}~\bibnamefont {Fu}},\
  }\href@noop {} {\bibfield  {journal} {\bibinfo  {journal} {Science advances}\
  }\textbf {\bibinfo {volume} {4}},\ \bibinfo {pages} {eaat2621} (\bibinfo
  {year} {2018})}\BibitemShut {NoStop}%
\bibitem [{\citenamefont {Kresse}\ and\ \citenamefont
  {Furthm\"uller}(1996{\natexlab{a}})}]{vasp1}%
  \BibitemOpen
  \bibfield  {author} {\bibinfo {author} {\bibfnamefont {G.}~\bibnamefont
  {Kresse}}\ and\ \bibinfo {author} {\bibfnamefont {J.}~\bibnamefont
  {Furthm\"uller}},\ }\href {\doibase 10.1103/PhysRevB.54.11169} {\bibfield
  {journal} {\bibinfo  {journal} {Phys. Rev. B}\ }\textbf {\bibinfo {volume}
  {54}},\ \bibinfo {pages} {11169} (\bibinfo {year}
  {1996}{\natexlab{a}})}\BibitemShut {NoStop}%
\bibitem [{\citenamefont {Kresse}\ and\ \citenamefont
  {Furthm\"uller}(1996{\natexlab{b}})}]{vasp2}%
  \BibitemOpen
  \bibfield  {author} {\bibinfo {author} {\bibfnamefont {G.}~\bibnamefont
  {Kresse}}\ and\ \bibinfo {author} {\bibfnamefont {J.}~\bibnamefont
  {Furthm\"uller}},\ }\href {\doibase
  https://doi.org/10.1016/0927-0256(96)00008-0} {\bibfield  {journal} {\bibinfo
   {journal} {Comput. Mater. Sci.}\ }\textbf {\bibinfo {volume} {6}},\ \bibinfo
  {pages} {15} (\bibinfo {year} {1996}{\natexlab{b}})}\BibitemShut {NoStop}%
\bibitem [{\citenamefont {Bl\"ochl}(1994)}]{PAW}%
  \BibitemOpen
  \bibfield  {author} {\bibinfo {author} {\bibfnamefont {P.~E.}\ \bibnamefont
  {Bl\"ochl}},\ }\href {\doibase 10.1103/PhysRevB.50.17953} {\bibfield
  {journal} {\bibinfo  {journal} {Phys. Rev. B}\ }\textbf {\bibinfo {volume}
  {50}},\ \bibinfo {pages} {17953} (\bibinfo {year} {1994})}\BibitemShut
  {NoStop}%
\bibitem [{\citenamefont {Perdew}\ \emph {et~al.}(1996)\citenamefont {Perdew},
  \citenamefont {Burke},\ and\ \citenamefont {Ernzerhof}}]{pbe}%
  \BibitemOpen
  \bibfield  {author} {\bibinfo {author} {\bibfnamefont {J.~P.}\ \bibnamefont
  {Perdew}}, \bibinfo {author} {\bibfnamefont {K.}~\bibnamefont {Burke}}, \
  and\ \bibinfo {author} {\bibfnamefont {M.}~\bibnamefont {Ernzerhof}},\ }\href
  {\doibase 10.1103/PhysRevLett.77.3865} {\bibfield  {journal} {\bibinfo
  {journal} {Phys. Rev. Lett.}\ }\textbf {\bibinfo {volume} {77}},\ \bibinfo
  {pages} {3865} (\bibinfo {year} {1996})}\BibitemShut {NoStop}%
\bibitem [{\citenamefont {Monkhorst}\ and\ \citenamefont
  {Pack}(1976)}]{MP_grid}%
  \BibitemOpen
  \bibfield  {author} {\bibinfo {author} {\bibfnamefont {H.~J.}\ \bibnamefont
  {Monkhorst}}\ and\ \bibinfo {author} {\bibfnamefont {J.~D.}\ \bibnamefont
  {Pack}},\ }\href {\doibase 10.1103/PhysRevB.13.5188} {\bibfield  {journal}
  {\bibinfo  {journal} {Phys. Rev. B}\ }\textbf {\bibinfo {volume} {13}},\
  \bibinfo {pages} {5188} (\bibinfo {year} {1976})}\BibitemShut {NoStop}%
\bibitem [{\citenamefont {Fu}\ and\ \citenamefont
  {Kane}(2007)}]{FuKane_parity}%
  \BibitemOpen
  \bibfield  {author} {\bibinfo {author} {\bibfnamefont {L.}~\bibnamefont
  {Fu}}\ and\ \bibinfo {author} {\bibfnamefont {C.~L.}\ \bibnamefont {Kane}},\
  }\href {\doibase 10.1103/PhysRevB.76.045302} {\bibfield  {journal} {\bibinfo
  {journal} {Phys. Rev. B}\ }\textbf {\bibinfo {volume} {76}},\ \bibinfo
  {pages} {045302} (\bibinfo {year} {2007})}\BibitemShut {NoStop}%
\end{thebibliography}%
\bibliographystyle{apsrev4-1}

\section{Acknowledgements}
This work was supported primarily by the U.S. Department of Energy, Office of Science, Basic Energy Sciences, under Award Number DE-SC0020149 (electrical measurements, tr-ARPES, and analysis) and also by the Gordon and Betty Moore Foundation?s EPiQS Initiative, Grant GBMF9070 to J.G.C. (materials synthesis and DFT calculations).


\section{Author Contributions}
J. Z. and T. S. synthesized the single crystals.  J. Z. performed the transport and torque measurements and analyzed the subsequent data.  C. L. and F. M. performed the tr-ARPES measurements and analyzed the subsequent data.  S. F. performed the density functional theory calculations.  J. Z. wrote the manuscript with input from all authors.  N. G. and J. G. C. coordinated the project.

\newpage


\newpage


\newpage



\newpage


\end{document}